\documentclass[11pt]{article}
\usepackage{amssymb,amsmath,amsfonts}
\usepackage{graphicx}
\usepackage{graphics}
\usepackage{eepic,epsfig}

1.60cm \makeatletter \@addtoreset{equation}{section}

\makeatother

\begin{document}

\title{Fermionic vacuum polarization in compactified \\
cosmic string spacetime }
\author{S. Bellucci$^{1}$\thanks{%
E-mail: bellucci@lnf.infn.it },\, E. R. Bezerra de Mello$^{2}$\thanks{%
E-mail: emello@fisica.ufpb.br},\, A. de Padua$^{2}$,\, A. A. Saharian$^{3}$
\thanks{%
E-mail: saharian@ysu.am} \\
\\
\textit{$^1$ INFN, Laboratori Nazionali di Frascati,}\\
\textit{Via Enrico Fermi 40, 00044 Frascati, Italy} \vspace{0.3cm}\\
\textit{$^{2}$Departamento de F\'{\i}sica, Universidade Federal da Para\'{\i}%
ba}\\
\textit{58.059-970, Caixa Postal 5.008, Jo\~{a}o Pessoa, PB, Brazil}\vspace{%
0.3cm}\\
\textit{$^3$Department of Physics, Yerevan State University,}\\
\textit{1 Alex Manoogian Street, 0025 Yerevan, Armenia}}
\maketitle

\begin{abstract}
We investigate the fermionic condensate (FC) and the vacuum expectation
value (VEV) of the energy-momentum tensor for a charged massive fermionic
field in the geometry of a cosmic string compactified along its axis. In
addition, we assume the presence of two types of magnetic fluxes: a flux
running along the cosmic string and another enclosed by the compact
dimension. These fluxes give rise to Aharanov-Bohm-like effects on the VEVs.
The VEVs are decomposed into two parts corresponding to the geometry of a
straight cosmic string without compactification plus a topological part
induced by the compactification of the string axis. Both contributions are
even periodic functions of the magnetic fluxes with period equal to the flux
quantum. The vacuum energy density is equal to the radial stress for the
parts corresponding to the straight cosmic string and the topological one.
Moreover, the axial stress is equal to the energy density for the parts
corresponding to the straight cosmic string; however, for massive fermionic
field this does not occur for the topological contributions. With respect to
the dependence on the magnetic fluxes, both, the fermionic condensate and
the vacuum energy density, can be either positive or negative. Moreover, for
points near the string, the main contribution to the VEVs comes from the
straight cosmic string part, whereas at large distances the topological ones
dominate. In addition to the local characteristics of the vacuum state, we
also evaluate the part in the topological Casimir energy induced by the
string.
\end{abstract}

\bigskip


\bigskip

\section{Introduction}

Topological and geometrical concepts are of great significance in the recent
developments of many areas of physics, including condensed matter physics,
gauge field theories and cosmology. In particular, the global properties of
the spacetime manifold play an important role in quantum field theory.
Several interesting quantum effects arise from the nontrivial topological
structure of background spacetime. A well-known example of such quantum
phenomena is the topological Casimir effect (for reviews see \cite{Eliz94}-%
\cite{Bord}). This effect is among the most striking macroscopic
manifestations of quantum properties of the vacuum state. The periodicity
conditions imposed on a quantum field due to nontrivial topology lead to a
modification of the spectrum of zero-point fluctuations and result in
shifting the vacuum expectation values (VEVs) for physical quantities. In
Kaluza-Klein type models, the dependence of the vacuum energy on the lengths
of extra dimensions can serve as a mechanism for the stabilization of moduli
fields. More recently, the topological Casimir effect has been also
considered as a model for dark energy driving the accelerated expansion of
the Universe \cite{Eliz01}-\cite{Zhi}. The topological Casimir effect in
cylindrical and toroidal carbon nanotubes is investigated in \cite%
{Bell09,Eliz11} within the framework of a Dirac-like theory for the
electronic states in graphene. In the present paper we investigate the
interplay of quantum topological effects for a charged massive fermionic
field coming from two different sources: from the conical geometry of a
cosmic string spacetime and from the compactification of the string axis.
Fermionic currents in this geometry, induced by magnetic fluxes, have been
recently studied in \cite{Saha13}. The topological Casimir densities for a
scalar field in compactified cosmic string spacetime are discussed in \cite%
{Mello12}. In this way, the present paper is a natural extension of the
investigations started in these references.

Cosmic strings are linear topological defects which play an important role
in cosmology. In the context of most unified particle physics models, these
objects may have been created by phase transitions in the very early
Universe \cite{V-S}. Topological defects of similar structure arise in a
number of condensed matter systems. Among the most important gravitational
effects of cosmic strings is the generation of a scale-invariant spectrum of
cosmological perturbations and, initially, the cosmic strings have been
considered as an alternative to inflation for generating primordial density
perturbations from which galaxies grew. Though the observational evidence
for acoustic oscillations in the angular power spectrum of cosmic microwave
background has ruled out cosmic strings as the dominant source for
primordial density perturbations, they are still candidates for the
generation of a number of interesting physical effects such as gamma ray
bursts \cite{Berezinski}, gravitational waves \cite{Damour} and high-energy
cosmic rays \cite{Bhattacharjee}. Moreover, the fundamental string theory
predicts the existence of macroscopic defects such as cosmic strings \cite%
{Sarangi}-\cite{Kibb04}.

At large distances from the cosmic string core, the spacetime geometry for
an infinite straight cosmic string has a conical topology with a planar
angle deficit proportional to the linear mass density. The vacuum
polarization effects in quantum field theory induced by this conical
structure have been considered in a large number of papers. In the specific
analysis for the VEV of the energy-momentum tensor, explicit calculations
have been developed associated with scalar, fermionic and electromagnetic
fields \cite{Hell86}-\cite{BezeKh06}. The Casimir-Polder forces acting on a
polarizable microparticle in the geometry of a cosmic string have been
investigated in \cite{Bard10}. For charged fields, considering the presence
of a magnetic flux running along the cosmic strings, there appear additional
contributions to the corresponding vacuum polarization effects \cite{charged}%
-\cite{Spin08}. The magnetic flux along the cosmic string induces also
vacuum current densities. This phenomenon has been investigated for scalar
fields in \cite{Sira,Yu}. The analysis of induced fermionic currents in
higher-dimensional cosmic string spacetime in the presence of a magnetic
flux have been developed in \cite{Mello10}. In these analysis the authors
have shown that induced vacuum current densities along the azimuthal
direction appear if the ratio of the magnetic flux by the quantum one has a
nonzero fractional part. Moreover, the fermionic current induced by a
magnetic flux in a $(2+1)$-dimensional conical spacetime and in the presence
of a circular boundary has also been analyzed in \cite{Saha10} (for the
combined effects of topology and boundaries on the quantum vacuum for
scalar, electromagnetic and fermionic fields in the geometry of a cosmic
string see \cite{Brev95}-\cite{Beze13b}).

The main objective of this paper is to investigate the combined effects of
planar angle deficit and of the compactification of cosmic string axis on
the fermionic condensate (FC) and on the VEV of the energy-momentum tensor.
We assume the presence of a magnetic flux running along the string axis and
the magnetic flux enclosed by the compact dimension. Although the
corresponding operators are local, due to the global nature of the vacuum
state, these VEVs carry important information about the global properties of
the background spacetime. The FC plays an important role in models of
dynamical breaking of chiral symmetry. The VEV of the energy-momentum tensor
acts as a source in the quasiclassical Einstein equations and is of key
importance in modelling self-consistent dynamics involving fermionic fields.

The paper is organized as follows. In the next section we present the
background geometry associated with the spacetime under consideration and
provide the complete set of normalized positive- and negative-energy
fermionic wave-functions obeying quasiperiodic boundary condition along the
string axis. By using the mode-summation procedure, we evaluate the FC. The
condensate is decomposed into two terms: the first one corresponds to the
geometry of a straight cosmic string with magnetic flux and the second term
is induced by the compactification of the string axis. The latter is an even
function of the magnetic fluxes, i.e., the fluxes along the string axis and
the one enclosed by the string axis. We have provided closed expressions for
both contributions. In Section \ref{sec3}, by using the mode-summation
procedure, we evaluate the VEVs for all components of the energy-momentum
tensor. Similar to the case of the FC, we provide a decomposition of these
VEVs into the sum of straight cosmic string and topological parts. Combined
expressions for the components of the vacuum energy-momentum tensor and
their asymptotics are presented in Section \ref{sec4}. In this sections we
also consider the part in the topological Casimir energy induced by the
cosmic string and the magnetic flux. The most relevant conclusions of the
paper are summarized in Section \ref{conc}. Throughout the paper we use the
units with $G=\hbar =c=1$

\section{Fermionic condensate}

\label{sec2}

\subsection{Mode functions}

In the presence of an external electromagnetic field with vector potential $%
A_{\mu }$, the quantum dynamic of a massive charged spinor field in curved
spacetime is governed by the Dirac equation,
\begin{equation}
i\gamma ^{\mu }{\mathcal{D}}_{\mu }\psi -m\psi =0\ ,\ {\mathcal{D}}_{\mu
}=\partial _{\mu }+\Gamma _{\mu }+ieA_{\mu },  \label{Direq}
\end{equation}%
where $\gamma ^{\mu }$ are the Dirac matrices in curved spacetime and $%
\Gamma _{\mu }$ is the spin connection. Both matrices are given in terms of
the flat spacetime Dirac matrices, $\gamma ^{(a)}$, by the relations,
\begin{equation}
\gamma ^{\mu }=e_{(a)}^{\mu }\gamma ^{(a)}\ ,\ \Gamma _{\mu }=\frac{1}{4}%
\gamma ^{(a)}\gamma ^{(b)}e_{(a)}^{\nu }e_{(b)\nu ;\mu }\ .  \label{Gammamu}
\end{equation}%
In (\ref{Gammamu}), $e_{(a)}^{\mu }$ represents the tetrad basis satisfying
the relation $e_{(a)}^{\mu }e_{(b)}^{\nu }\eta ^{ab}=g^{\mu \nu }$, with $%
\eta ^{ab}$ being the Minkowski spacetime metric tensor.

The four dimensional spacetime corresponding to an idealized cosmic string
along the $z$-axis, can be written, by using cylindrical coordinates,
through the line element below:
\begin{equation}
ds^{2}=dt^{2}-dr^{2}-r^{2}d\phi ^{2}-dz{}^{2}\ .  \label{ds21}
\end{equation}%
Here the coordinates take values in the ranges $r\geqslant 0$, $0\leqslant
\phi \leqslant \phi _{0}=2\pi /q$, $-\infty <t<+\infty $. The parameter $q$,
bigger than unity, is related to the linear mass density of the string, $\mu
_{0}$, by $q^{-1}=1-4\mu _{0}$. In the geometry described by (\ref{ds21})
the gamma matrices can be taken in the form \cite{Saha13}
\begin{equation}
\gamma ^{0}=\gamma ^{(0)}=\left(
\begin{array}{cc}
1 & 0 \\
0 & -1%
\end{array}%
\right) ,\;\gamma ^{l}=\left(
\begin{array}{cc}
0 & \sigma ^{l} \\
-\sigma ^{l} & 0%
\end{array}%
\right) \ ,  \label{gamcurved}
\end{equation}%
where for the $2\times 2$ matrices $\sigma ^{l}$, $l=(r,\ \phi ,\ z)$, one
has
\begin{equation}
\sigma ^{r}=\left(
\begin{array}{cc}
0 & e^{-iq\phi } \\
e^{iq\phi } & 0%
\end{array}%
\right) \ ,\ \sigma ^{\phi }=-\frac{i}{r}\left(
\begin{array}{cc}
0 & e^{-iq\phi } \\
-e^{iq\phi } & 0%
\end{array}%
\right) \ ,\ \sigma ^{z}=\left(
\begin{array}{cc}
1 & 0 \\
0 & -1%
\end{array}%
\right) \ .  \label{betl}
\end{equation}%
It is easy to check that with this choice the matrices (\ref{gamcurved})
obey the Clifford algebra with the metric tensor from (\ref{ds21}).

In the analysis that we want to develop, it will be assumed that the
direction along the $z$-axis is compactified to a circle with length $L$: $%
0\leqslant z\leqslant L$. Along the compact dimension we impose the
quasiperiodicity condition,
\begin{equation}
\psi (t,r,\phi ,z+L)=e^{2\pi i\beta }\psi (t,r,\phi ,z)\ ,  \label{Period}
\end{equation}%
with a constant phase $\beta $, $0\leqslant \beta \leqslant 1$. In addition,
we shall admit the existence of a gauge field with the constant vector
potential
\begin{equation}
A_{\mu }=(0,0,A_{\phi },A_{z})\ .  \label{Amu}
\end{equation}%
The component $A_{\phi }$ is related to an infinitesimal thin magnetic flux,
$\Phi _{\phi }$, running along the string by $A_{\phi }=-q\Phi _{\phi
}/(2\pi )$ (note that $A_{\phi }$ and $A_{z}$ are the covariant components
of the 4-vector $A_{\mu }=(0,-\mathbf{A})$ with $\mathbf{A}$ being the
corresponding 3-vector). Similarly, the axial component $A_{z}$ can be given
in terms of the magnetic flux $\Phi _{z}$ enclosed by the $z$-axis as $%
A_{z}=-\Phi _{z}/L$. Though the magnetic field strength corresponding to (%
\ref{Amu}) vanishes, the nontrivial topology of the background geometry
leads to Aharonov-Bohm-like effects on the VEVs of physical observables.

In the present paper we are interested in the effects of the string
compactification along its axis on the fermionic condensate (FC) and on the
VEV of the energy-momentum tensor. For the evaluation of these VEVs, a
complete set of fermionic mode-functions is needed. In Ref. \cite{Saha13},
we have shown that the positive- and negative-energy fermionic
mode-functions are uniquely specified by the set of quantum number $\sigma
=(\lambda ,k,j,s)$. These functions can be written in the form
\begin{equation}
\psi _{\sigma }^{(\pm )}(x)=C_{\sigma }^{(\pm )}e^{\mp iEt+kz+iq(j-1/2)\phi
}\left(
\begin{array}{c}
J_{\beta _{j}}(\lambda r) \\
sJ_{\beta _{j}+\epsilon _{j}}(\lambda r)e^{iq\phi } \\
\pm \frac{\tilde{k}_{l}-is\epsilon _{j}\lambda }{E\pm m}J_{\beta
_{j}}(\lambda r) \\
\mp s\frac{\tilde{k}_{l}-is\lambda \epsilon _{j}}{E\pm m}J_{\beta
_{j}+\epsilon _{j}}(\lambda r)e^{iq\phi }%
\end{array}%
\right) \ ,  \label{psi+n}
\end{equation}%
where $J_{\nu }(x)$ is the Bessel function, $s=\pm 1$, $\lambda \geq 0$, and
\begin{equation}
\beta _{j}=q|j+\alpha |-\epsilon _{j}/2\ ,\;\alpha =eA_{\phi }/q=-\Phi
_{\phi }/\Phi _{0},  \label{betaj}
\end{equation}%
with $\epsilon _{j}=\mathrm{sgn}(j+\alpha )$ and with $\Phi _{0}=2\pi /e$
being the flux quantum. The mode-functions (\ref{psi+n}) are eigenfunctions
for the projection of total angular momentum operator along the cosmic
string,
\begin{equation}
\widehat{J}_{3}\psi _{\sigma }^{(\pm )}=\left( -i\partial _{\phi }+i\frac{q}{%
2}\gamma ^{(1)}\gamma ^{(2)}\right) \psi _{\sigma }^{(\pm )}=qj\psi _{\sigma
}^{(\pm )}\ ,  \label{J3}
\end{equation}%
with eigenvalues $j=\pm 1/2,\pm 3/2,\ldots $. The eigenvalues of the axial
quantum number $k$ are determined by the periodicity condition (\ref{Period}%
),
\begin{equation}
k=k_{l}=2\pi (l+\beta )/L\ ,\ \ l=0,\pm 1,\pm 2\ ,\ldots \ .  \label{Eigkz}
\end{equation}%
The energy is expressed in terms of $\lambda $ and $l$ by the relation
\begin{equation}
E=\sqrt{\lambda ^{2}+\tilde{k}_{l}^{2}+m^{2}}\ ,\;\tilde{k}_{l}=2\pi (l+%
\tilde{\beta})/L,  \label{E+}
\end{equation}%
where%
\begin{equation}
\tilde{\beta}=\beta +eA_{z}L/(2\pi )=\beta -\Phi _{z}/\Phi _{0}\ .
\label{bett}
\end{equation}%
Note that, in order to simplify presentation of the mode-functions, in the
negative-energy modes we have changed the signs of the quantum numbers $%
(k,j,s)$, compared with Ref. \cite{Saha13}.

The constants $C_{\sigma }^{(\pm )}$ in (\ref{psi+n}) are determined from
the orthonormalization condition
\begin{equation}
\int d^{3}x\sqrt{\gamma }\ (\psi _{\sigma }^{(r)})^{\dagger }\psi _{\sigma
^{\prime }}^{(r^{\prime })}=\delta _{\sigma \sigma ^{\prime }}\ \delta
_{rr^{\prime }},\;r,r^{\prime }=+,-,  \label{normcond}
\end{equation}%
where $\gamma $ is the determinant of the spatial metric tensor. The delta
symbol on the right-hand side is understood as the Dirac delta function for
continuous quantum numbers ($\lambda $) and the Kronecker delta for discrete
ones ($k,j,s,r$). From (\ref{normcond}) one finds%
\begin{equation}
|C_{\sigma }^{(\pm )}|^{2}=\frac{q\lambda (E\pm m)}{8\pi LE}\ .  \label{C+}
\end{equation}%
In deriving the mode-functions (\ref{psi+n}) we have imposed regularity
condition on the string axis. A discussion on the contribution from
irregular modes is given in \cite{Saha13}.

The VEVs of physical observables will depend on $\beta $ and $\Phi _{z}$ in
the combination given by (\ref{bett}). This result could be seen directly by
making use of the gauge \ transformation $A_{\mu }=A_{\mu }^{\prime
}+\partial _{\mu }\Lambda (x)$, $\psi (x)=\psi ^{\prime }(x)e^{-ie\Lambda
(x)}$ with the function $\Lambda (x)=A_{z}z$. The new function $\psi
^{\prime }(x)$ obeys the Dirac equation with $A_{z}^{\prime }=0$ and the
periodicity condition $\psi ^{\prime }(t,r,\phi ,z+L)=e^{2\pi i\tilde{\beta}%
}\psi ^{\prime }(t,r,\phi ,z)$. The VEVs are not changed under this gauge
transformation and in the new gauge a single parameter $\tilde{\beta}$
appears instead of $\beta $ and $A_{z}$.

Having the complete set of wave-functions we are in condition to evaluate
the FC induced by the compactification of the string along its axis and also
by the magnetic fluxes. The FC is defined as the VEV $\langle 0|\bar{\psi}%
\psi |0\rangle \equiv \langle \bar{\psi}\psi \rangle $, where $|0\rangle $
corresponds to the vacuum state, and $\bar{\psi}=\psi ^{\dagger }\gamma
^{(0)}$ is the Dirac adjoint. Expanding the field operator in terms of the
complete set $\{\psi _{\sigma }^{(+)},\ \psi _{\sigma }^{(-)}\}$ and by
using the standard anticommutation relations for annihilation and creation
operators, the following formula for the FC is obtained:
\begin{equation}
\langle \bar{\psi}\psi \rangle =\sum_{\sigma }\bar{\psi}_{\sigma }^{(-)}\psi
_{\sigma }^{(-)}\ ,  \label{FCmodesum}
\end{equation}%
where we use the compact notation defined as
\begin{equation}
\sum_{\sigma }=\int_{0}^{\infty }d\lambda \ \sum_{l=-\infty }^{+\infty
}\sum_{s=\pm 1}\sum_{j=\pm 1/2,\cdots }\ .  \label{Sumsig}
\end{equation}%
This VEV\ is a periodic function of the fluxes $\Phi _{\phi }$ and $\Phi
_{z} $ with the period equal to the flux quantum. In particular, if we write
the parameter $\alpha $ in (\ref{betaj}) in the form%
\begin{equation}
\alpha =n_{0}+\alpha _{0},\;|\alpha _{0}|<1/2,  \label{alfa0}
\end{equation}%
where $n_{0}$ is an integer number, the FC will depend on $\alpha _{0}$
only. Note that, for the boundary condition at the cone apex used in \cite%
{Saha10}, there are no square integrable irregular modes for $|\alpha
_{0}|\leqslant (1-1/q)/2$.

Substituting the wave-function (\ref{psi+n}) into (\ref{FCmodesum}), we can
see that the terms with $s=1$ and $s=-1$ give the same contribution and one
gets
\begin{equation}
\langle \bar{\psi}\psi \rangle =-\frac{qm}{2\pi L}\sum_{l=-\infty }^{+\infty
}\sum_{j}\int_{0}^{\infty }d\lambda \,\frac{\lambda }{E}[J_{\beta
_{j}}^{2}(\lambda r)+J_{\beta _{j}+\epsilon _{j}}^{2}(\lambda r)]\ .
\label{FC1}
\end{equation}%
In what follows we shall use the notation
\begin{equation}
\sum_{j}=\sum_{j=\pm 1/2,\cdots } \ .  \label{SumNot}
\end{equation}%
Of course, the expression in the right-hand side of (\ref{FC1}) is divergent
and a regularization is necessary. Here we assume the presence of a cutoff
function without writing it explicitly. As we shall see, the specific form
of this function is not relevant in the discussion below. By taking into
account the expression (\ref{E+}) for the energy, the summation over the
quantum number $l$ can be developed by using the Abel-Plana summation
formula in the form \cite{Bell10},\footnote{%
For generalizations of the Abel-Plana formula see \cite{SahaBook}.}%
\begin{equation}
\frac{2\pi }{L}\sum_{l=-\infty }^{\infty }f(|\tilde{k}_{l}|)=2\int_{0}^{%
\infty }dk\,f(k)+i\int_{0}^{\infty }dk\sum_{\delta =\pm 1}\frac{f(ik)-f(-ik)%
}{e^{Lk+2\pi i\delta \tilde{\beta}}-1}\ ,  \label{SumForm}
\end{equation}%
taking $f(k)=(k^{2}+\lambda ^{2}+m^{2})^{-1/2}$. Consequently the FC can be
written in the decomposed form:
\begin{equation}
\langle \bar{\psi}\psi \rangle =\langle \bar{\psi}\psi \rangle _{s}+\langle
\bar{\psi}\psi \rangle _{c}\ ,  \label{FCdec}
\end{equation}%
where $\langle \bar{\psi}\psi \rangle _{s}$ is the contribution due to the
first integral in the right-hand side of (\ref{SumForm}) and corresponds to
the FC in the geometry of cosmic string spacetime in the absence of
compactification. As to the term $\langle \bar{\psi}\psi \rangle _{c}$, it
vanishes in the limit $L\rightarrow \infty $ and this contribution is
induced by the compactification of the string along its axis.

In the absence of the magnetic flux, corresponding to $\alpha _{0}=0$, a
closed analytic expression for $\langle \bar{\psi}\psi \rangle _{s}$ is
provided in \cite{Beze13b}. In this special case an alternative integral
representation is derived in \cite{Tar08}. The FC in a $(2+1)$-dimensional
conical spacetime in the presence of a magnetic flux, has been evaluated in
\cite{Bell11} for a massive field obeying MIT bag boundary condition on a
circular boundary. To our knowledge, a closed expression for the FC in a
four-dimensional cosmic string spacetime in the presence of magnetic flux
running along the string has not been obtained yet. So in order to fulfill
this blank, we shall include this calculation in this paper.

\subsection{FC in the geometry of straight cosmic string}

Combining (\ref{FC1}) and (\ref{SumForm}), for the FC in the geometry of a
straight cosmic string we get the integral representation
\begin{eqnarray}
\langle \bar{\psi}\psi \rangle _{s} &=&-\frac{qm}{2\pi ^{2}}\int_{0}^{\infty
}d\lambda \lambda \int_{0}^{\infty }dk\frac{1}{\sqrt{k^{2}+\lambda ^{2}+m^{2}%
}}  \notag \\
&&\times \sum_{j}[J_{\beta _{j}}^{2}(\lambda r)+J_{\beta _{j}+\epsilon
_{j}}^{2}(\lambda r)]\ .  \label{FC2}
\end{eqnarray}%
For the further transformation, we use the relation
\begin{equation}
\frac{1}{\sqrt{k^{2}+\lambda ^{2}+m^{2}}}=\frac{2}{\sqrt{\pi }}%
\int_{0}^{\infty }ds\ e^{-(k^{2}+\lambda ^{2}+m^{2})s^{2}}\ .  \label{ident1}
\end{equation}%
Substituting this into (\ref{FC2}), we can easily integrate over the
variable $k$. As to the integral over $\lambda $, we use the integral
involving the square of the Bessel function from \cite{Grad} with the result
\begin{equation}
\int_{0}^{\infty }d\lambda \,\lambda e^{-s^{2}\lambda ^{2}}\left[ J_{\beta
_{j}}^{2}(\lambda r)+J_{\beta _{j}+\epsilon _{j}}^{2}(\lambda r)\right] =%
\frac{e^{-y}}{2s^{2}}\left[ I_{\beta _{j}}(y)+I_{\beta _{j}+\epsilon _{j}}(y)%
\right] \ ,  \label{Int-reg}
\end{equation}%
with $y=r^{2}/(2s^{2})$ and with $I_{\nu }(z)$ being the modified Bessel
function. As a result, the FC is presented in the form%
\begin{equation}
\langle \bar{\psi}\psi \rangle _{s}=-\frac{qm}{(2\pi r)^{2}}\int_{0}^{\infty
}dy\ e^{-y-m^{2}r^{2}/(2y)}\mathcal{J}(q,\alpha _{0},y),  \label{FC3}
\end{equation}%
where we have defined the function%
\begin{equation}
\mathcal{J}(q,\alpha _{0},y)=\mathcal{I}(q,\alpha _{0},y)+\mathcal{I}%
(q,-\alpha _{0},y),  \label{JCal}
\end{equation}%
with $\mathcal{I}(q,\alpha _{0},y)=\sum_{j}I_{\beta _{j}}(y)$ and $\mathcal{I%
}(q,-\alpha _{0},y)=\sum_{j}I_{\beta _{j}+\epsilon _{j}}(y)$.

An integral representation for the function $\mathcal{I}(q,\alpha _{0},y)$,
suitable for the extraction of the divergent part in the FC, is derived in
\cite{Saha10}. By using that representation, for the function (\ref{JCal})
one finds the following formula:
\begin{eqnarray}
\mathcal{J}(q,\alpha _{0},y) &=&\frac{2}{q}e^{y}+\frac{4}{\pi }%
\int_{0}^{\infty }dx\,\frac{h(q,\alpha _{0},x)\sinh x}{\cosh (2qx)-\cos
(q\pi )}e^{-y\cosh (2x)}  \notag \\
&&+\frac{4}{q}\sum_{k=1}^{p}(-1)^{k}\cos (\pi k/q)\cos (2\pi k\alpha
_{0})e^{y\cos (2\pi k/q)}\ ,  \label{Sum01}
\end{eqnarray}%
where $p$ is an integer defined by $2p\leqslant q<2p+2$ and for $1\leqslant
q<2$ the last term on the right-hand side is absent. The function in the
integrand of (\ref{Sum01}) is given by the expression
\begin{eqnarray}
h(q,\alpha _{0},x) &=&\cos \left[ q\pi \left( 1/2+\alpha _{0}\right) \right]
\sinh \left[ \left( 1-2\alpha _{0}\right) qx\right]  \notag \\
&&+\cos \left[ q\pi \left( 1/2-\alpha _{0}\right) \right] \sinh \left[
\left( 1+2\alpha _{0}\right) qx\right] \ .  \label{g0}
\end{eqnarray}%
Note that $\mathcal{J}(q,\alpha _{0},y)$ is an even function of $\alpha _{0}$%
.

In the case of integer values of $q$ and for
\begin{equation}
\alpha _{0}=\frac{1}{2}-\frac{n+1/2}{q}\ ,  \label{gammmaSp}
\end{equation}%
with an integer $n$, one has $h(q,\alpha _{0},x)=0$. From the condition $%
|\alpha _{0}|<1/2$ we find $0\leqslant n<q-1/2$. In this case a simpler
expression for the function $\mathcal{J}(q,\alpha _{0},y)$ is obtained:
\begin{equation}
\mathcal{J}(q,\alpha _{0},y)=\frac{2}{q}\sum_{k=0}^{q-1}\cos (\pi k/q)\cos
((2n+1)\pi k/q)e^{y\cos (2\pi k/q)}.  \label{ISp}
\end{equation}

We can see that the first term on the right-hand side of (\ref{Sum01})
provides a contribution to FC independent of $\alpha_{0}$ and $q$. It
corresponds to the FC in Minkowski spacetime in the absence of magnetic
flux. This term provides a divergent result to FC. Because the geometry of
the cosmic string is flat outside the string core, the renormalization for $%
\langle \bar{\psi}\psi \rangle _{s}$ reduces to subtract from this
expression the Minkowski spacetime part. So we discard the exponential term
in \eqref{Sum01}. The other terms provide contributions to the FC due to the
magnetic flux and nontrivial topology of the straight cosmic string. These
terms are finite and do not require any renormalization procedure. So the
cutoff function, assumed implicitly before, can be safely removed.
Substituting (\ref{Sum01}) into (\ref{FC3}), the integrals over the variable
$y$ are evaluated with the help of formula from \cite{Grad}, and the final
result for the renormalized FC is written as:
\begin{eqnarray}
\langle \bar{\psi}\psi \rangle _{s}^{\mathrm{ren}} &=&-\frac{2m^{3}}{\pi ^{2}%
}\left[ \sum_{k=1}^{p}(-1)^{k}\cos (\pi k/q)\cos (2\pi k\alpha
_{0})f_{1}(2mrs_{k})\right.  \notag \\
&&+\left. \frac{q}{\pi }\int_{0}^{\infty }dx\frac{h(q,\alpha _{0},x)\sinh x}{%
\cosh (2qx)-\cos (q\pi )}f_{1}(2mr\cosh x)\right] \ .  \label{FC4}
\end{eqnarray}%
Here we have introduced the notations
\begin{equation}
f_{\nu }(x)=K_{\nu }(x)/x^{\nu },\;s_{k}=\sin (\pi k/q),  \label{sk}
\end{equation}%
with $K_{\nu }(z)$ being the Macdonald function. In the absence of the
magnetic flux one has $\alpha _{0}=0$ and, hence, $h(q,0,x)=2\cos (q\pi
/2)\sinh (qx)$. In this case the formula (\ref{FC4}) is reduced to the one
derived in \cite{Beze13b}. For $q=1$, Eq. (\ref{FC4}) gives the FC induced
by the magnetic flux in Minkowski spacetime. For the special case mentioned
in (\ref{gammmaSp}), the integral term vanishes and the renormalized value
for the FC is expressed by,
\begin{equation}
\langle \bar{\psi}\psi \rangle _{s}^{\mathrm{ren}}=-\frac{m^{3}}{\pi ^{2}}%
\sum_{k=1}^{q-1}\cos (\pi k/q)\cos ((2n+1)\pi k/q)f_{1}(2mrs_{k})\ .
\label{FCSpec}
\end{equation}

Let us consider some limiting cases. First of all, by using the asymptotic%
\begin{equation}
f_{\nu }(x)\sim 2^{\nu -1}\Gamma (\nu )x^{-2\nu },\;x\rightarrow 0,
\label{fnuas}
\end{equation}%
we see that the FC vanishes for a massless field. At large distances from
the string, $mr\gg 1$, the FC is suppressed by the factor $e^{-2mr}$ for $%
1\leqslant q\leqslant 2$ and by the factor $e^{-2mr\sin (\pi /q)}$ for $q>2$%
. And finally, on the string, $r\rightarrow 0$, the FC diverges as $1/r^{2}$.

\begin{figure}[tbph]
\begin{center}
\epsfig{figure=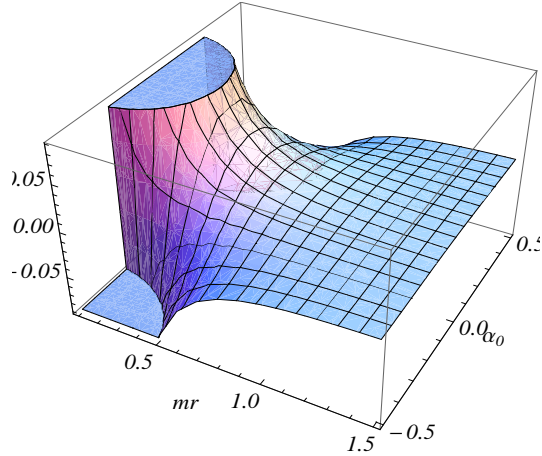,width=7.cm,height=6.cm}
\end{center}
\caption{FC in the geometry of a straight cosmic string, $\langle \bar{%
\protect\psi}\protect\psi \rangle _{s}^{\mathrm{ren}}/m^{3}$, versus the
distance from the string and the parameter $\protect\alpha _{0}$. For the
parameter describing the planar angle deficit we have taken $q=2.5$.}
\label{fig1}
\end{figure}

In figure \ref{fig1} we have plotted the FC in the geometry of a straight
cosmic string, $\langle \bar{\psi}\psi \rangle _{s}^{\mathrm{ren}}/m^{3}$,
as a function of the distance from the string and of the parameter $\alpha
_{0}$ characterizing the magnetic flux along the string. For the parameter
describing the planar angle deficit we have taken the value $q=2.5$. As we
see, in dependence of the flux, the FC can be either positive or negative.
In the absence of the magnetic flux the FC is positive.

\subsection{Topological part}

Let us now develop the calculation for the contribution to the FC induced by
the compactification. This part comes from the second integral in (\ref%
{SumForm}) and is presented in the form
\begin{eqnarray}
\langle \bar{\psi}\psi \rangle _{c} &=&-\frac{qm}{2\pi ^{2}}%
\sum_{j}\int_{0}^{\infty }d\lambda \,\lambda \int_{\sqrt{\lambda ^{2}+m^{2}}%
}^{\infty }dk  \notag \\
&&\times \frac{J_{\beta _{j}}^{2}(\lambda r)+J_{\beta _{j}+\epsilon
_{j}}^{2}(\lambda r)}{\sqrt{k^{2}-\lambda ^{2}-m^{2}}}\sum_{\delta =\pm 1}%
\frac{1}{e^{Lk+2\delta \pi i\tilde{\beta}}-1}\ .  \label{FCc1}
\end{eqnarray}%
To continue the calculation, in the integrand of (\ref{FCc1}) we use the
series expansion%
\begin{equation}
\left( e^{u}-1\right) ^{-1}=\sum_{l=1}^{\infty }e^{-lu}.  \label{Exp}
\end{equation}%
The integral over $k$ is expressed in terms of the function $K_{0}(x)$ and
the above expression becomes
\begin{equation}
\langle \bar{\psi}\psi \rangle _{c}=-\frac{qm}{\pi ^{2}}\sum_{l=1}^{\infty
}\cos (2\pi l\tilde{\beta})\int_{0}^{\infty }d\lambda \lambda K_{0}(lL\sqrt{%
\lambda ^{2}+m^{2}})\sum_{j}[J_{\beta _{j}}^{2}(\lambda r)+J_{\beta
_{j}+\epsilon _{j}}^{2}(\lambda r)]\ .  \label{FCc3}
\end{equation}

By making use of the integral representation for the Macdonald function,
\begin{equation}
K_{\nu }(x)=\frac{1}{2}\left( \frac{x}{2}\right) ^{\nu }\int_{0}^{\infty }dt%
\frac{e^{-t-x^{2}/(4t)}}{t^{\nu +1}}\ ,  \label{Kintergral}
\end{equation}%
we see that the integral over $\lambda $ becomes of the form (\ref{Int-reg}%
). Defining a new variable $y=2tr^{2}/(l^{2}L^{2})$, the above expression is
written as,
\begin{eqnarray}
\langle \bar{\psi}\psi \rangle _{c} &=&-\frac{qm}{2\pi ^{2}r^{2}}%
\sum_{l=1}^{\infty }\cos (2\pi l\tilde{\beta})\int_{0}^{\infty }dy\   \notag
\\
&&\times e^{-y[1+l^{2}L^{2}/(2r^{2})]-m^{2}r^{2}/(2y)}\mathcal{J}(q,\alpha
_{0},y)\ .  \label{FCc5}
\end{eqnarray}%
The integrand in this expression is nonnegative and, hence, the topological
part in the FC is always negative for $\tilde{\beta}=0$. In this case $%
|\langle \bar{\psi}\psi \rangle _{c}|$ is a monotonically decreasing
function of $L$. Substituting the expression (\ref{Sum01}) into (\ref{FCc5})
and integrating over $y$ we get the expression:
\begin{eqnarray}
\langle \bar{\psi}\psi \rangle _{c} &=&-\frac{4m^{3}}{\pi ^{2}}%
\sum_{l=1}^{\infty }\cos (2\pi l\tilde{\beta})\left[ \sideset{}{'}{\sum}%
_{k=0}^{p}(-1)^{k}\cos (\pi k/q)\cos (2\pi k\alpha _{0})f_{1}(mL\sqrt{%
l^{2}+\rho _{k}^{2}})\right.  \notag \\
&&+\left. \frac{q}{\pi }\int_{0}^{\infty }dx\frac{h(q,\alpha _{0},x)\sinh x}{%
\cosh (2qx)-\cos (q\pi )}f_{1}(mL\sqrt{l^{2}+\eta ^{2}(x)})\right] \ ,
\label{FCc6}
\end{eqnarray}%
where we have introduced the notations
\begin{equation}
\rho _{k}=\frac{2r}{L}\sin (\pi k/q),\;\eta (x)=\frac{2r}{L}\cosh x.
\label{rho-eta}
\end{equation}%
In (\ref{FCc6}), the prime on the summation over $k$ means that the term
with $k=0$ should be taken with the weight 1/2. For a massless field the
topological part vanishes. In the special case defined by (\ref{gammmaSp}),
the topological part in the FC reads:
\begin{eqnarray}
\langle \bar{\psi}\psi \rangle _{c} &=&-\frac{2m^{3}}{\pi ^{2}}%
\sum_{l=1}^{\infty }\cos (2\pi l\tilde{\beta})\sum_{k=0}^{q-1}\cos \left(
\pi k/q\right)  \notag \\
&&\times \cos (\left( 2n+1\right) \pi k/q)f_{1}(mL\sqrt{l^{2}+\rho _{k}^{2}}%
)\ .  \label{FCSp}
\end{eqnarray}

In the absence of cosmic string and flux along the $z$-axis one has $q=1$, $%
\alpha _{0}=0$, and in (\ref{FCc6}) the term $k=0$ survives only with the
result
\begin{equation}
\langle \bar{\psi}\psi \rangle _{c}^{(0)}=-\frac{2m^{3}}{\pi ^{2}}%
\sum_{l=1}^{\infty }\cos (2\pi l\tilde{\beta})f_{1}(mLl)\ .  \label{FCM}
\end{equation}%
This quantity corresponds to the FC in Minkowski spacetime with spatial
topology $R^{2}\times S^{1}$ and with the length of the compact dimension $L$%
. The FC for a more general case of spatial topology $R^{p}\times
(S^{1})^{D-p}$ has been investigated in \cite{Bell09}. The topological part (%
\ref{FCc6}) is finite on the string:
\begin{equation}
\langle \bar{\psi}\psi \rangle _{c}=[1+2s_{0}(q,\alpha _{0})]\langle \bar{%
\psi}\psi \rangle _{c}^{(0)},\;r=0,  \label{FCc7}
\end{equation}%
with $\langle \bar{\psi}\psi \rangle _{c}^{(0)}$ given by (\ref{FCM}) and
with the notation%
\begin{eqnarray}
s_{n}(q,\alpha _{0}) &=&\sum_{k=1}^{p}\frac{(-1)^{k}}{s_{k}^{n}}{\cos (}\pi
k/q{)\cos (}2\pi k\alpha _{0}{)}  \notag \\
&&+\frac{q}{\pi }\int_{0}^{\infty }dx\frac{h(q,\alpha _{0},x)\sinh x\,\cosh
^{-n}x}{\cosh (2qx)-\cos (q\pi )}.  \label{sq}
\end{eqnarray}%
Hence, near the string the the FC is dominated by the part $\langle \bar{\psi%
}\psi \rangle _{s}$. In the absence of the magnetic flux, for the function (%
\ref{sq}) we have:%
\begin{eqnarray}
s_{0}(q,0) &=&-\frac{1}{2},\;s_{2}(q,0)=\frac{1-q^{2}}{12},  \notag \\
s_{4}(q,0) &=&\frac{1-q^{2}}{720}(7q^{2}+17).  \label{s024}
\end{eqnarray}%
Now from (\ref{FCc7}) we see that in the absence of the magnetic flux the
topological part in the FC vanishes on the string axis.

Combining the expressions (\ref{FC4}) and (\ref{FCc6}), the total FC is
written in the form%
\begin{eqnarray}
\langle \bar{\psi}\psi \rangle &=&\langle \bar{\psi}\psi \rangle _{c}^{(0)}-%
\frac{4m^{3}}{\pi ^{2}}\sideset{}{'}{\sum}_{l=0}^{\infty }\cos (2\pi l\tilde{%
\beta})\left[ \sum_{k=1}^{p}(-1)^{k}\cos (\pi k/q)\cos (2\pi k\alpha
_{0})f_{1}(mL\sqrt{l^{2}+\rho _{k}^{2}})\right.  \notag \\
&&\left. +\frac{q}{\pi }\int_{0}^{\infty }dx\frac{h(q,\alpha _{0},x)\sinh x}{%
\cosh (2qx)-\cos (q\pi )}f_{1}(mL\sqrt{l^{2}+\eta ^{2}(x)})\right] \ .
\label{FCtot}
\end{eqnarray}%
Here also, the prime means that the term with $l=0$ should be taken with the
factor $1/2$. The latter presents the part $\langle \bar{\psi}\psi \rangle
_{s}^{\mathrm{ren}}$. The second term in the right-hand side of this formula
encodes the effects from the string and from the flux running along its
axis. At large distances from the string, $mr\gg 1$, these effects are
suppressed by the factor $e^{-2mr\sin (\pi /q)}$ for $q>2$ and by the factor
$e^{-2mr}$ for $1\leqslant q\leqslant 2$, and one has $\langle \bar{\psi}%
\psi \rangle \approx \langle \bar{\psi}\psi \rangle _{c}^{(0)}$.

In the limit $mL\ll 1$, the asymptotic behavior of the topological part in
the FC depends crucially on the parameter $\tilde{\beta}$, if it is zero or
not. For $mL\ll 1$, the dominant contribution in the second term of the
right-hand side in (\ref{FCtot}) comes from large values of $l$ and we can
replace the summation over $l$ by the integration. By using the integration
formula from \cite{Grad}, for the corresponding integral we find%
\begin{equation}
\int_{0}^{\infty }dl\cos (al)f_{1}(\sqrt{l^{2}c^{2}+b^{2}})=\frac{\pi }{2bc}%
e^{-b\sqrt{\left( a/c\right) ^{2}+1}}.  \label{IntForm4}
\end{equation}%
For $\tilde{\beta}=0$ this gives%
\begin{equation}
\langle \bar{\psi}\psi \rangle \approx \langle \bar{\psi}\psi \rangle
_{c}^{(0)}+\frac{2}{L}\langle \bar{\psi}\psi \rangle _{s}^{(D=2)}.
\label{FCsmalL}
\end{equation}%
where%
\begin{eqnarray}
\langle \bar{\psi}\psi \rangle _{s}^{(D=2)} &=&-\frac{m}{2\pi r}\left[
\sum_{k=1}^{p}\frac{(-1)^{k}\cos (\pi k/q)}{s_{k}e^{2mrs_{k}}}\cos (2\pi
k\alpha _{0})\right.  \notag \\
&&\left. +\frac{q}{\pi }\int_{0}^{\infty }dx\frac{h(q,\alpha _{0},x)\sinh x}{%
\cosh (2qx)-\cos (q\pi )}\frac{e^{-2mr\cosh x}}{\cosh x}\right] ,
\label{FCD2}
\end{eqnarray}%
is the FC in a $(2+1)$-dimensional conical spacetime ($D=2$) \cite{Bell11}.
An additional coefficient 2 in (\ref{FCsmalL}) is related to the fact that
the number of spinor components in $D=2$ is 2 instead of 4-component spinors
in $D=3$. For $\tilde{\beta}\neq 0$, in (\ref{FCtot}) the part induced by
the string is suppressed by the factor $\exp [-4\pi \tilde{\beta}r\sin (\pi
/q)/L]$ for $q>2$ and by the factor $\exp (-4\pi \tilde{\beta}r/L)$ for $%
1\leqslant q\leqslant 2$.

\begin{figure}[tbph]
\begin{center}
\epsfig{figure=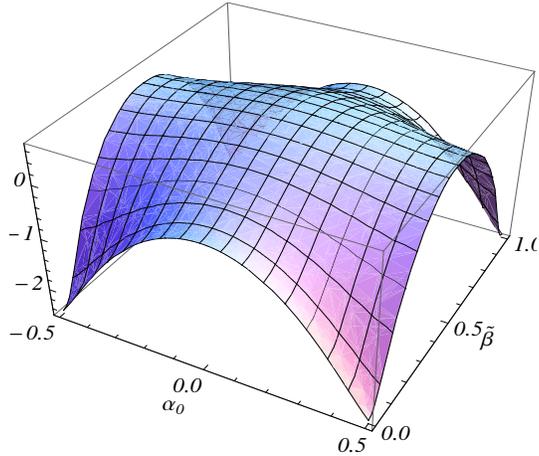,width=7.cm,height=6.cm}
\end{center}
\caption{FC as a function of the parameters $\protect\alpha _{0}$ and $%
\tilde{\protect\beta}$ for $mr=0.25$, $mL=0.5$ in the geometry of cosmic
string with $q=2.5$.}
\label{fig2}
\end{figure}

The figure \ref{fig2} presents the quantity $\langle \bar{\psi}\psi \rangle
/m^{3}$ as a function of the parameters $\alpha _{0}$ and $\tilde{\beta}$
for $mr=0.25$, $mL=0.5$ in the geometry of cosmic string with $q=2.5$. As to
the previous analysis, the dependence of FC on the parameters can provide
positive or negative values.

\begin{figure}[tbph]
\begin{center}
\epsfig{figure=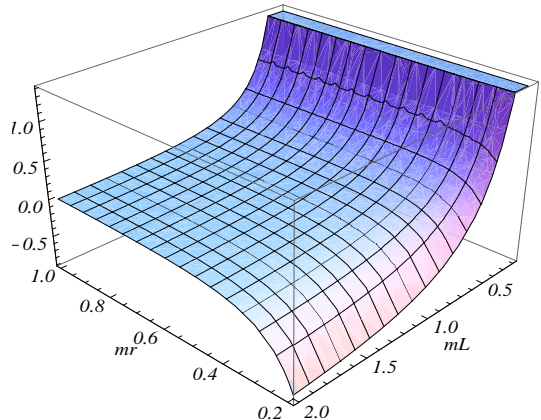,width=7.cm,height=6.cm}
\end{center}
\caption{FC as a function of the distance from the string and on the length
of compactification for $\protect\alpha _{0}=\tilde{\protect\beta}=0.5$ and $%
q=2.5$. }
\label{fig3}
\end{figure}

The dependence of the FC on the distance from the string and on the length
of compactification is displayed in figure \ref{fig3} for special values $%
\alpha _{0}=\tilde{\beta}=0.5$ and for $q=2.5$.

\section{Energy-momentum tensor}

\label{sec3}

In this section we analyze another important characteristic of the fermionic
vacuum, the VEV of the energy-momentum tensor. For a charged fermionic
field, in the presence of electromagnetic field, the operator of the
energy-momentum tensor is expressed as:
\begin{equation}
T_{\mu \nu }=\frac{i}{2}\left[ \bar{\psi}\gamma _{(\mu }{\mathcal{D}}_{\nu
)}\psi -({\mathcal{D}}_{(\mu }\bar{\psi})\gamma _{\nu )}\psi \right] \,.
\label{EMTdef}
\end{equation}%
where ${\mathcal{D}}_{\mu }\bar{\psi}=\partial _{\mu }\bar{\psi}-ieA_{\mu }%
\bar{\psi}-\bar{\psi}\Gamma _{\mu }$ and the brackets in the index
expression mean the symmetrization over the enclosed indices. Similar to the
case of the FC, the VEV of the energy-momentum tensor, $\langle 0|T_{\mu \nu
}|0\rangle \equiv \langle T_{\mu \nu }\rangle $, can be evaluated by using
the mode-sum formula
\begin{equation}
\langle T_{\mu \nu }\rangle =\frac{i}{2}\sum_{\sigma }\left[ \bar{\psi}%
_{\sigma }^{(-)}\gamma _{(\mu }{\mathcal{D}}_{\nu )}\psi _{\sigma }^{(-)}-({%
\mathcal{D}}_{(\mu }\bar{\psi}_{\sigma })\gamma _{\nu )}\psi _{\sigma }^{(-)}%
\right] \,,  \label{EMT}
\end{equation}%
with the notation (\ref{Sumsig}). As in the case of the FC, we assume the
presence of a cutoff function. We shall evaluate separately all components
of the energy-momentum tensor.

In the geometry under consideration, the VEV of the energy-momentum tensor
can be decomposed as%
\begin{equation}
\langle T_{\mu }^{\nu }\rangle =\langle T_{\mu }^{\nu }\rangle _{s}+\langle
T_{\mu }^{\nu }\rangle _{c},  \label{TmunuDec}
\end{equation}%
where the term $\langle T_{\mu }^{\nu }\rangle _{s}$ is the VEV for the
geometry of an infinite straight cosmic string, and $\langle T_{\mu }^{\nu
}\rangle _{c}$ is the contribution coming from the compactification of the
string along its axis. An important point to be mentioned here is that the
compactification does not change the local geometry and, hence, the
divergences in the VEVs of $\langle T_{\mu }^{\nu }\rangle $ and $\langle
T_{\mu }^{\nu }\rangle _{s}$ are the same. Thus, by the decomposition (\ref%
{TmunuDec}), the renormalization of $\langle T_{\mu }^{\nu }\rangle $ is
reduced to the one for $\langle T_{\mu }^{\nu }\rangle _{s}$. As in the case
of the FC, an explicit decomposition of the form (\ref{TmunuDec}) can be
obtained by using the summation formula (\ref{SumForm}).

\subsection{Energy density}

Let us first consider the energy density, $\langle T_{0}^{0}\rangle $. By
taking into account that $A_{0}$ and $\Gamma _{0}$ vanish and $\partial
_{t}\psi _{\sigma }^{(-)}=iE\psi _{\sigma }^{(-)}$, we can see that the
contributions of the terms $s=1$ and $s=-1$ are the same. After the
summation over $s$ the mode-sum (\ref{EMT}) for the energy density is
reduced to
\begin{equation}
\langle T_{0}^{0}\rangle =-\frac{q}{2\pi L}\sum_{j}\sum_{l=-\infty
}^{+\infty }\int_{0}^{\infty }d\lambda \,\lambda E[J_{\beta
_{j}}^{2}(\lambda r)+J_{\beta _{j}+\epsilon _{j}}^{2}(\lambda r)]\ .
\label{T001}
\end{equation}%
The summation over the quantum number $l$ can be developed by using the
formula (\ref{SumForm}) with $f(k)=(k^{2}+\lambda ^{2}+m^{2})^{1/2}$. The
parts in the VEV of the energy density $\langle T_{0}^{0}\rangle _{s}$ and $%
\langle T_{0}^{0}\rangle _{c}$ correspond to the first and second terms in
the right-hand side of (\ref{SumForm}), respectively.

We start with the evaluation of $\langle T_{0}^{0}\rangle _{s}$. It is
written in the form
\begin{equation}
\langle T_{0}^{0}\rangle _{s}=-\frac{q}{2\pi ^{2}}\sum_{j}\int_{0}^{\infty
}d\lambda \,\lambda \int_{0}^{\infty }dk\ \sqrt{k^{2}+\lambda ^{2}+m^{2}}%
[J_{\beta _{j}}^{2}(\lambda r)+J_{\beta _{j}+\epsilon _{j}}^{2}(\lambda r)]\
.  \label{T002}
\end{equation}%
To present this expression in a form more suitable for the renormalization
and for numerical calculations of the renormalized part, we use the relation
\begin{equation}
\sqrt{k^{2}+\lambda ^{2}+m^{2}}=-\frac{2}{\sqrt{\pi }}\int_{0}^{\infty
}ds\,\partial _{s^{2}}e^{-(k^{2}+\lambda ^{2}+m^{2})s^{2}}\ .  \label{Ident2}
\end{equation}%
After the substitution of (\ref{Ident2}) into (\ref{T002}), the integration
over $k$ is elementary and the integral over $\lambda $ is evaluated by
using (\ref{Int-reg}). As a result, the VEV of the energy density is written
as:
\begin{equation}
\langle T_{0}^{0}\rangle _{s}=-\frac{q}{2\pi ^{2}r^{4}}\int_{0}^{\infty }dy\
y^{1/2}\partial _{y}[y^{3/2}e^{-y-m^{2}r^{2}/(2y)}\mathcal{J}(q,\alpha
_{0},y)]\ ,  \label{T004}
\end{equation}%
where the expression for $\mathcal{J}(q,\alpha _{0},y)$ is given by (\ref%
{Sum01}). The first term in the right-hand side of (\ref{Sum01}), $2e^{y}/q$%
, provides the divergent part of the VEV. This contribution corresponds to
the VEV of the energy density in Minkowski spacetime and in the absence of
magnetic fluxes. The well defined physical result for the energy density is
obtained by applying the renormalization procedure, which is reduced to the
subtraction of the Minkoswskian counterpart.

So, by making use of (\ref{Sum01}), after the integration over $y$, the
renormalized energy density is expressed as:
\begin{eqnarray}
\langle T_{0}^{0}\rangle _{s}^{\mathrm{ren}} &=&\frac{2m^{4}}{\pi ^{2}}\left[
\sum_{k=1}^{p}(-1)^{k}{\cos }(\pi k/q){\cos }(2\pi k\alpha
_{0})f_{2}(2mrs_{k})\right.  \notag \\
&+&\left. \frac{q}{\pi }\int_{0}^{\infty }dx\frac{h(q,\alpha _{0},x)\sinh x}{%
\cosh (2qx)-\cos (q\pi )}f_{2}(2mr\cosh x)\right] \ ,  \label{T005}
\end{eqnarray}%
with the notation (\ref{sk}). For the special case defined in (\ref{gammmaSp}%
), this expression is reduced to
\begin{equation}
\langle T_{0}^{0}\rangle _{s}^{\mathrm{ren}}=\frac{m^{4}}{\pi ^{2}}%
\sum_{k=1}^{q-1}\cos (\pi k/q)\cos ((2n+1)\pi k/q)f_{2}(2mrs_{k})\ .
\label{T00Sp}
\end{equation}

Now we turn to the part in the VEV of the energy density induced by the
compactification. This part is given by the second integral in the
right-hand side of (\ref{SumForm}). By using the expansion (\ref{Exp}),
after the integration over $k$ we can write the topological part in the form
\begin{eqnarray}
\langle T_{0}^{0}\rangle _{c} &=&\frac{q}{\pi ^{2}L}\sum_{l=1}^{\infty }%
\frac{\cos (2\pi l\tilde{\beta})}{l}\sum_{j}\int_{0}^{\infty }d\lambda
\,\lambda \sqrt{\lambda ^{2}+m^{2}}  \notag \\
&&\times K_{1}(lL\sqrt{\lambda ^{2}+m^{2}})[J_{\beta _{j}}^{2}(\lambda
r)+J_{\beta _{j}+\epsilon _{j}}^{2}(\lambda r)]\ .  \label{T00c2}
\end{eqnarray}%
By taking into account that, $K_{1}(x)=-K_{0}^{\prime }(x)$ and using the
integral representation (\ref{Kintergral}), the integral over $\lambda $ is
evaluated with the help of (\ref{Int-reg}). This leads to the
representation:
\begin{eqnarray}
\langle T_{0}^{0}\rangle _{c} &=&\frac{q}{\pi ^{2}r^{2}}\sum_{l=1}^{\infty }%
\frac{\cos (2\pi l\tilde{\beta})}{l^{2}L^{2}}\int_{0}^{\infty
}dy\,e^{-yl^{2}L^{2}/(2r^{2})-r^{2}m^{2}/(2y)}  \notag \\
&&\times \left( \partial _{y}+\frac{r^{2}m^{2}}{2y^{2}}\right) ye^{-y}%
\mathcal{J}(q,\alpha _{0},y).  \label{T00c2b}
\end{eqnarray}%
In the part of the integral with $\partial _{y}$ we integrate by parts with
the result:%
\begin{equation}
\langle T_{0}^{0}\rangle _{c}=\frac{q}{2\pi ^{2}r^{4}}\sum_{l=1}^{\infty
}\cos (2\pi l\tilde{\beta})\int_{0}^{\infty
}dy\,ye^{-y[1+l^{2}L^{2}/(2r^{2})]-r^{2}m^{2}/(2y)}\mathcal{J}(q,\alpha
_{0},y).  \label{T00c2c}
\end{equation}%
From here it follows that for $\tilde{\beta}=0$ the topological part in the
VEV of the energy density is a monotonically decreasing positive function
with respect to the both $L$ and $m$. Now, by using (\ref{Sum01}), after the
integration over $y$ we arrive to the final expression
\begin{eqnarray}
\langle T_{0}^{0}\rangle _{c} &=&\frac{4m^{4}}{\pi ^{2}}\sum_{l=1}^{\infty }{%
\cos }(2\pi l\tilde{\beta})\left[ \sideset{}{'}{\sum}_{k=0}^{p}(-1)^{k}\cos
(\pi k/q)\cos (2\pi k\alpha _{0})f_{2}(mL\sqrt{l^{2}+\rho _{k}^{2}})\right.
\notag \\
&+&\left. \frac{q}{\pi }\int_{0}^{\infty }dx\frac{h(q,\alpha _{0},x)\sinh x}{%
\cosh (2qx)-\cos (q\pi )}f_{2}(mL\sqrt{l^{2}+\eta ^{2}(x)})\right] \ ,
\label{T00c4}
\end{eqnarray}%
where $\rho _{k}$ and $\eta (z)$ are defined in (\ref{rho-eta}). In the
special case defined in \eqref{gammmaSp}, we obtain,
\begin{equation}
\langle T_{0}^{0}\rangle _{c}=\frac{2m^{4}}{\pi ^{2}}\sum_{l=1}^{\infty }{%
\cos (2\pi l{\tilde{\beta}})}\sum_{k=0}^{q-1}\cos \left[ (2\pi k/q)(n+1/2)%
\right] \cos (\pi k/q)f_{2}(mL\sqrt{l^{2}+\rho _{k}^{2}})\ .
\end{equation}

\subsection{Radial stress}

Our next step is the evaluation of the radial stress, $\langle
T_{r}^{r}\rangle $. In order to do that, we take $A_{r}=\Gamma _{r}=0$ in
the general definition of the covariant derivative of the fermionic field.
In this way, we can write,
\begin{equation}
\langle T_{r}^{r}\rangle =\frac{i}{2}\sum_{\sigma }\left[ \bar{\psi}_{\sigma
}^{(-)}\gamma ^{r}(\partial _{r}\psi _{\sigma }^{(-)})-(\partial _{r}\bar{%
\psi}_{\sigma }^{(-)})\gamma ^{r}\psi _{\sigma }^{(-)}\right] \ .
\label{TrrModeSum}
\end{equation}%
Substituting the mode-functions from (\ref{psi+n}) into the above
expression, after some intermediate steps, we arrive at,
\begin{equation}
\langle T_{r}^{r}\rangle =-\frac{q}{4\pi L}\sum_{\sigma }\frac{\epsilon
_{j}\lambda ^{3}}{E}[J_{\beta _{j}}^{\prime }(\lambda r)J_{\beta
_{j}+\epsilon _{j}}(\lambda r)-J_{\beta _{j}}(\lambda r)J_{\beta
_{j}+\epsilon _{j}}^{\prime }(\lambda r)]\ ,  \label{Trr1}
\end{equation}%
where the primes means derivative with respect to the argument of the
function. By using the recurrent relations for the Bessel functions, after
the summation over $s$, we write the VEV in the form
\begin{equation}
\langle T_{r}^{r}\rangle =\frac{q}{2\pi L}\sum_{l=-\infty }^{+\infty
}\int_{0}^{\infty }d\lambda \,\frac{\lambda ^{3}}{E}S(\lambda r),
\label{Trr11}
\end{equation}%
with the function%
\begin{equation}
S(x)=\sum_{j}\left[ J_{\beta _{j}}^{2}(x)+J_{\beta _{j}+\epsilon
_{j}}^{2}(x)-\frac{2\beta _{j}+\epsilon _{j}}{x}J_{\beta _{j}}(x)J_{\beta
_{j}+\epsilon _{j}}(x)\right] .  \label{S1}
\end{equation}%
After the application of the summation formula (\ref{SumForm}) to the series
over $l$ in (\ref{Trr11}), the radial stress is decomposed into the parts $%
\langle T_{r}^{r}\rangle _{s}$ and $\langle T_{r}^{r}\rangle _{c}$.

We start with the calculation of the part corresponding to the geometry of a
straight cosmic string:
\begin{equation}
\langle T_{r}^{r}\rangle _{s}=\frac{2q}{(2\pi )^{2}}\int_{0}^{\infty }\
d\lambda \,\lambda ^{3}\int_{0}^{\infty }dk\frac{S(\lambda r)}{\sqrt{%
k^{2}+\lambda ^{2}+m^{2}}}\ .  \label{Trr2}
\end{equation}%
Using the relation (\ref{ident1}) we can develop the integration over $k$.
After that we can write the above expression in the form:
\begin{equation}
\langle T_{r}^{r}\rangle _{s}=\frac{2q}{(2\pi )^{2}}\int_{0}^{\infty }ds\,%
\frac{e^{-m^{2}s^{2}}}{s}\int_{0}^{\infty }d\lambda \,\lambda
^{3}e^{-\lambda ^{2}s^{2}}S(\lambda r)\ .
\end{equation}%
Compared with the case of the energy density, the integral over $\lambda $
is more delicate. Below we present the main steps:
\begin{eqnarray}
\int_{0}^{\infty }d\lambda \,\lambda ^{3}{e^{-\lambda ^{2}s^{2}}}S(\lambda
r) &=&\sum_{j}\int_{0}^{\infty }d\lambda \,\lambda ^{3}e^{-\lambda ^{2}s^{2}}%
\left[ J_{\beta _{j}}^{2}(\lambda r)+J_{\beta _{j}+\epsilon
_{j}}^{2}(\lambda r)\right]  \notag \\
&&-\sum_{j}\frac{2\beta _{j}+\epsilon _{j}}{r}\int_{0}^{\infty }d\lambda
\,\lambda ^{2}e^{-\lambda ^{2}s^{2}}J_{\beta _{j}}(\lambda r)J_{\beta
_{j}+\epsilon _{j}}(\lambda r)\ .  \label{calc1}
\end{eqnarray}%
The first integral on the right-hand side of (\ref{calc1}) can be developed
taking the derivative $-\partial _{s^{2}}$ on the integral in the left-hand
side of (\ref{Int-reg}). As to the second integral, by using the relation%
\begin{equation*}
J_{\beta _{j}}(\lambda r)J_{\beta _{j}+\epsilon _{j}}(\lambda r)=\frac{1}{%
2\lambda }\left( -\epsilon _{j}\partial _{r}+2\beta _{j}/r\right) J_{\beta
_{j}}^{2}(\lambda r),
\end{equation*}%
we can show that,
\begin{equation}
\int_{0}^{\infty }d\lambda {\lambda ^{2}}{e^{-\lambda ^{2}s^{2}}}\ J_{\beta
_{j}}(\lambda r)J_{\beta _{j}+\epsilon _{j}}(\lambda r)=\frac{r\epsilon
_{j}e^{-y}}{4s^{4}}[I_{\beta _{j}}(y)-I_{\beta _{j}+\epsilon _{j}}(y)]\ ,
\label{ident-1}
\end{equation}%
with $y=r^{2}/(2s^{2})$. Further, we use the relation
\begin{equation}
(1+2\epsilon _{j}\beta _{j})[I_{\beta _{j}}(y)-I_{\beta _{j}+\epsilon
_{j}}(y)]=2\left( y\partial _{y}-y+1/2\right) [I_{\beta _{j}}(y)+I_{\beta
_{j}+\epsilon _{j}}(y]\ ,  \label{IdentD}
\end{equation}%
to express the term on the right-hand side of (\ref{ident-1}) as the sum of
the modified Bessel functions. Combining all these results, for the VEV of
the radial stress we obtain:
\begin{equation}
\langle T_{r}^{r}\rangle _{s}=\frac{q}{4\pi ^{2}r^{4}}\int_{0}^{\infty
}dy\,ye^{-y-m^{2}r^{2}/(2y)}\mathcal{J}(q,\alpha _{0},y)]\ .  \label{Trrcs2}
\end{equation}%
By the direct substitution of (\ref{Sum01}) into (\ref{Trrcs2}), we can see
that the divergent part coming from the first term in the right-hand side of
(\ref{Sum01}) presents the corresponding quantity on topologically trivial
Minkowski spacetime in the absence of fluxes. Applying again the standard
renormalization procedure, we can see that the renormalized expression for
the radial stress has the same structure as that obtained from (\ref{T004})
after the integration by parts. Consequently, we conclude that%
\begin{equation}
\langle T_{r}^{r}\rangle _{s}^{\mathrm{ren}}=\langle T_{0}^{0}\rangle _{s}^{%
\mathrm{ren}} \ .  \label{TrrT00}
\end{equation}

Now we pass to the evaluation of the part $\langle T_{r}^{r}\rangle _{c}$.
This contribution comes from the second integral on the right-hand side of
the summation formula (\ref{SumForm}). By using the expansion (\ref{Exp}),
the integral over $k$ is expressed in terms of the function $K_{0}(lL\sqrt{%
m^{2}+\lambda ^{2}})$. With the help of the integral representation (\ref%
{Kintergral}) for the Macdonald function, the expression for the topological
part becomes,
\begin{equation}
\langle T_{r}^{r}\rangle _{c}=\frac{q}{2\pi ^{2}}\sum_{l=1}^{\infty }\cos
(2\pi l\tilde{\beta})\int_{0}^{\infty }dt\frac{e^{-t-l^{2}L^{2}m^{2}/(4t)}}{t%
}\int_{0}^{\infty }d\lambda {\lambda ^{3}}{e^{-l^{2}L^{2}\lambda ^{2}/(4t)}}%
S(\lambda r)\ .  \label{Trrc2}
\end{equation}%
The integral over $\lambda $ is in the form (\ref{calc1}). So its result is:
\begin{equation}
\int_{0}^{\infty }d\lambda {\lambda ^{3}}{e^{-l^{2}L^{2}\lambda ^{2}/(4t)}}%
S(\lambda r)\ \ =\frac{y^{2}}{r^{4}}e^{-y}\sum_{j}[I_{\beta
_{j}}(y)+I_{\beta _{j}+\epsilon _{j}}(y)]\ ,  \label{IntRelS}
\end{equation}%
with the notation $y=2r^{2}t/(l^{2}L^{2})$. Finally we arrive to the
expression which coincides with (\ref{T00c2c}). Hence, for the topological
parts as well we have the relation%
\begin{equation}
\langle T_{r}^{r}\rangle _{c}=\langle T_{0}^{0}\rangle _{c}\ .  \label{Tcrr}
\end{equation}%
Note that, for a scalar field with general curvature coupling parameter, the
VEVs of the radial stress and the energy density are different \cite{Mello12}%
.

\subsection{Azimuthal stress}

In the evaluation of the VEV for the azimuthal stress, $\langle T_{\phi
}^{\phi }\rangle $, we have to take into account, $A_{\phi }=q\alpha /e$ and
\begin{equation}
\Gamma _{\phi }=\frac{1-q}{2}\gamma ^{(1)}\gamma ^{(2)}=-\frac{i}{2}%
(1-q)\Sigma ^{(3)}\ ,\ \Sigma ^{(3)}=\mathrm{diag}(\sigma _{3},\sigma _{3})\
,  \label{Gam}
\end{equation}%
being $\sigma _{3}$ the Pauli matrix. So this component reads,
\begin{equation}
\langle T_{\phi }^{\phi }\rangle =\frac{i}{2}\sum_{\sigma }[\bar{\psi}%
_{\sigma }^{(-)}\gamma ^{\phi }{\mathcal{D}}_{\phi }\psi _{\sigma }^{(-)}-({%
\mathcal{D}}_{\phi }\bar{\psi}_{\sigma }^{(-)})\gamma ^{\phi }\psi _{\sigma
}^{(-)}]\ .  \label{Tphi}
\end{equation}%
In the development of the term inside the bracket, it is convenient to
express the angular derivative in terms of the total angular momentum
operator: $\partial _{\phi }=i\widehat{J}_{3}-i(q/2)\Sigma ^{(3)}$.
Moreover, we can observe that the anticommutator, $\{\gamma ^{\phi },\Sigma
^{(3)}\}$, which appears in the development, vanishes. So after some steps,
we get:
\begin{equation}
\langle T_{\phi }^{\phi }\rangle =q\sum_{\sigma }(j+\alpha )\bar{\psi}%
_{\sigma }^{(-)}\gamma ^{\phi }\psi _{\sigma }^{(-)}\ .  \label{Tpp}
\end{equation}%
Now substituting the Dirac matrix $\gamma ^{\phi }$ and the expression for
the negative-energy wave-function, we obtain
\begin{equation}
\langle T_{\phi }^{\phi }\rangle =\frac{q^{2}}{2\pi Lr}\sum_{\sigma
}\epsilon _{j}(j+\alpha )\frac{\lambda ^{2}}{E}J_{\beta _{j}}(\lambda
r)J_{\beta _{j}+\epsilon _{j}}(\lambda r)\ .  \label{Tpp1}
\end{equation}%
The summation over $s$ provides the factor $2$ and for the summation over $l$
we use again (\ref{SumForm}). In this way, we have for the azimuthal stress
the decomposition (\ref{TmunuDec}).

Let us start with the part corresponding to the straight cosmic string:
\begin{equation}
\langle T_{\phi }^{\phi }\rangle _{s}=\frac{q^{2}}{\pi ^{2}r}%
\sum_{j}\epsilon _{j}(j+\alpha )\int_{0}^{\infty }d\lambda \,\lambda
^{2}\int_{0}^{\infty }dk\frac{J_{\beta _{j}}(\lambda r)J_{\beta
_{j}+\epsilon _{j}}(\lambda r)}{\sqrt{k^{2}+\lambda ^{2}+m^{2}}}\ .
\label{Tphi2}
\end{equation}%
Using again the relation (\ref{ident1}), we can develop the integral over $k$%
. As to the integral over $\lambda $ we use (\ref{ident-1}). Finally
defining the new variable $y=r^{2}/(2s^{2})$ we get,
\begin{equation}
\langle T_{\phi }^{\phi }\rangle _{s}=\frac{q^{2}}{2\pi ^{2}r^{4}}%
\sum_{j}(j+\alpha )\int_{0}^{\infty }dy\,ye^{-y-m^{2}r^{2}/(2y)}\left[
I_{\beta _{j}}(y)-I_{\beta _{j}+\epsilon _{j}}(y)\right] \ .  \label{Tppcs1}
\end{equation}%
By taking into account that $q(j+\alpha )=\epsilon _{j}\beta _{j}+1/2$ and
using the relation (\ref{IdentD}), we present (\ref{Tppcs1}) in the form%
\begin{equation}
\langle T_{\phi }^{\phi }\rangle _{s}=\frac{q}{2\pi ^{2}r^{4}}%
\int_{0}^{\infty }dy\,ye^{-m^{2}r^{2}/(2y)}\left( y\partial _{y}+1/2\right)
e^{-y}\mathcal{J}(q,\alpha _{0},y)].  \label{Tphi3}
\end{equation}%
Comparing this expression with (\ref{Trrcs2}) one can see that the following
relation takes place:%
\begin{equation}
\langle T_{\phi }^{\phi }\rangle _{s}=\left( r\partial _{r}+1\right) \langle
T_{r}^{r}\rangle _{s}.  \label{Tphir}
\end{equation}

Combining (\ref{Tphir}) with the expression for $\langle T_{r}^{r}\rangle
_{s}$ and by using the relations
\begin{equation}
f_{\nu }^{\prime }(x)=-xf_{\nu +1}(x),\;x^{2}f_{\nu +1}(x)=2\nu f_{\nu
}(x)+f_{\nu -1}(x),  \label{relfnu}
\end{equation}%
for the renormalized VEV of the azimuthal stress we find \ the following
result:
\begin{eqnarray}
\langle T_{\phi }^{\phi }\rangle _{s}^{\mathrm{ren}} &=&\frac{2m^{4}}{\pi
^{2}}\left[ \sum_{k=1}^{p}(-1)^{k}\cos (\pi k/q)\cos (2\pi k\alpha
_{0})F_{\phi }^{(0)}(2mrs_{k})\right.  \notag \\
&&\left. +\frac{q}{\pi }\int_{0}^{\infty }dx\frac{h(q,\alpha _{0},x)\sinh x}{%
\cosh (2qx)-\cos (q\pi )}F_{\phi }^{(0)}(2mr\cosh x)\right] \ ,
\label{Tppcs2}
\end{eqnarray}%
with the notation%
\begin{equation}
F_{\phi }^{(0)}(x)=\partial _{x}\left[ xf_{2}(x)\right]
=f_{2}(x)-x^{2}f_{3}(x)=-f_{1}(x)-3f_{2}(x).  \label{Fphi}
\end{equation}

Now we start the evaluation of the contribution to the azimuthal stress due
to the compactification. This term is given by the substitution of the
second integral in (\ref{SumForm}) into (\ref{Tpp1}). By using the expansion
(\ref{Exp}), the integral over $k$ provides $K_{0}(lL\sqrt{m^{2}+\lambda ^{2}%
})$. On the base of the integral representation (\ref{Kintergral}) we can
rewrite the above expression as:
\begin{eqnarray}
\langle T_{\phi }^{\phi }\rangle _{c} &=&\frac{q^{2}}{\pi ^{2}r}%
\sum_{l=1}^{\infty }\cos (2\pi l\tilde{\beta})\int_{0}^{\infty }dt\,\frac{%
e^{-t-l^{2}L^{2}m^{2}/(4t)}}{t}\sum_{j}\epsilon _{j}(j+\alpha )  \notag \\
&&\times \int_{0}^{\infty }d\lambda \,\lambda ^{2}e^{-l^{2}L^{2}\lambda
^{2}/(4t)}J_{\beta _{j}}(\lambda r)J_{\beta _{j}+\epsilon _{j}}(\lambda r)\ .
\label{Tppc2}
\end{eqnarray}%
The integral over $\lambda $ can be obtained by using the previous result (%
\ref{ident-1}). Defining a new variable $y=2r^{2}t/(l^{2}L^{2})$, the
expression obtained is given in terms of $q(j+\alpha )\left[ I_{\beta
_{j}}(y)-I_{\beta _{j}+\epsilon _{j}}(y)\right] $. To continue the
development, we use again (\ref{IdentD}) and get,
\begin{eqnarray}
\langle T_{\phi }^{\phi }\rangle _{c} &=&\frac{q}{\pi ^{2}r^{4}}%
\sum_{l=1}^{\infty }\cos (2\pi l\tilde{\beta})\int_{0}^{\infty }dy\ y\
e^{-y[1+l^{2}L^{2}/(2r^{2})]-m^{2}r^{2}/(2y)}  \notag \\
&&\times (y\partial _{y}-y+1/2)\mathcal{J}(q,\alpha _{0},y)]\ .
\label{Tppc3}
\end{eqnarray}%
By using the same trick as for the case of $\langle T_{\phi }^{\phi }\rangle
_{s}$, we can see that from (\ref{Tppc3}) the following relation is obtained:%
\begin{equation}
\langle T_{\phi }^{\phi }\rangle _{c}=\left( r\partial _{r}+1\right) \langle
T_{r}^{r}\rangle _{c}.  \label{Tphirc}
\end{equation}%
Now combining this with the expression for $\langle T_{r}^{r}\rangle _{c}$,
one gets the final expression for the topological part in the azimuthal
stress:%
\begin{eqnarray}
\langle T_{\phi }^{\phi }\rangle _{c} &=&\frac{4m^{4}}{\pi ^{2}}%
\sum_{l=1}^{\infty }{\cos }(2\pi l\tilde{\beta})\left[ \sideset{}{'}{\sum}%
_{k=0}^{p}(-1)^{k}\cos (\pi k/q)\cos (2\pi k\alpha _{0})F_{\phi }^{(l)}(mL%
\sqrt{l^{2}+\rho _{k}^{2}})\right.  \notag \\
&&+\left. \frac{q}{\pi }\int_{0}^{\infty }dx\frac{h(q,\alpha _{0},x)\sinh x}{%
\cosh (2qx)-\cos (q\pi )}F_{\phi }^{(l)}(mL\sqrt{l^{2}+\eta ^{2}(x)})\right]
\ ,  \label{Tppc4}
\end{eqnarray}%
with the function%
\begin{equation}
F_{\phi }^{(l)}(x)=\left( m^{2}L^{2}l^{2}-x^{2}\right) f_{3}(x)+f_{2}(x).
\label{Fphil}
\end{equation}%
Note that for $l=0$ this function coincides with (\ref{Fphi}).

\subsection{Axial stress}

In the calculation of the axial stress, we have to consider $A_{z}=-\Phi
_{z}/L$ in the covariant derivative of the field operator. So we have, ${%
\mathcal{D}}_{z}\psi _{\sigma }^{(-)}=i\tilde{k}_{l}\psi _{\sigma }^{(-)}$.
In addition, the matrix $\gamma ^{z}$ coincides with the standard expression
for the Dirac matrix in flat spacetime. For this component, we have,
\begin{equation}
\langle T_{z}^{z}\rangle =-\sum_{\sigma }\tilde{k}_{l}\bar{\psi}_{\sigma
}^{(-)}\gamma ^{z}\psi _{\sigma }^{(-)}\ .  \label{TzzMode}
\end{equation}%
Substituting the expression for the negative-energy mode function into the
above expression, one obtains
\begin{equation}
\langle T_{z}^{z}\rangle =-\frac{q}{2\pi L}\sum_{j}\int_{0}^{\infty
}d\lambda \,\lambda \sum_{l=-\infty }^{+\infty }\frac{\tilde{k}_{l}^{2}}{E}%
[J_{\beta _{j}}^{2}(\lambda r)+J_{\beta _{j}+\epsilon _{j}}^{2}(\lambda r)]\
.  \label{Tzz1}
\end{equation}%
For the summation over $l$, we use the Abel-Plana summation formula, Eq. (%
\ref{SumForm}), taking $f(k)=k^{2}(k^{2}+\lambda ^{2}+m^{2})^{-1/2}$. This
allows us to decompose the axial stress in accordance with (\ref{TmunuDec}).
The contribution to the axial stress corresponding to the geometry of a
straight cosmic string is given by the first term on the right-hand side of (%
\ref{SumForm}):
\begin{equation}
\langle T_{z}^{z}\rangle _{s}=\frac{q}{2\pi ^{2}}\sum_{j}\int_{0}^{\infty
}d\lambda \,\lambda \int_{0}^{\infty }dk\,k^{2}\frac{J_{\beta
_{j}}^{2}(\lambda r)+J_{\beta _{j}+\epsilon _{j}}^{2}(\lambda r)}{\sqrt{%
m^{2}+\lambda ^{2}+k^{2}}}\ .  \label{Tzzcs1}
\end{equation}%
Using the relation (\ref{ident1}), the evaluation of the integral over $k$
can be promptly obtained. As to the integral over $\lambda $ we use again
the result (\ref{Int-reg}). Finally, we get,
\begin{equation}
\langle T_{z}^{z}\rangle _{s}=\frac{q}{4\pi ^{2}r^{4}}\int_{0}^{\infty }dy\
y\ e^{-y-m^{2}r^{2}/(2y)}\mathcal{J}(q,\alpha _{0},y)]\ .  \label{Tzzcs2a}
\end{equation}%
This result coincides with (\ref{Trrcs2}), consequently we conclude that:
\begin{equation}
\langle T_{z}^{z}\rangle _{s}^{\mathrm{ren}}=\langle T_{0}^{0}\rangle _{s}^{%
\mathrm{ren}}\ .
\end{equation}%
Also, this property directly follows from the invariance of the problem with
respect to the boost along the axis of the string.

The topological part in the axial stress is given by substituting the second
integral on the right-hand side of (\ref{SumForm}) into (\ref{Tzz1}):
\begin{eqnarray}
\langle T_{z}^{z}\rangle _{c} &=&-\frac{q}{\pi ^{2}}\sum_{l=1}^{\infty }\cos
(2\pi l\tilde{\beta})\int_{0}^{\infty }d\lambda \,\lambda \int_{\sqrt{%
\lambda ^{2}+m^{2}}}^{\infty }dk\,  \notag \\
&&\times \frac{k^{2}e^{-lLk}}{\sqrt{k^{2}-\lambda ^{2}-m^{2}}}%
\sum_{j}[J_{\beta _{j}}^{2}(\lambda r)+J_{\beta _{j}+\epsilon
_{j}}^{2}(\lambda r)]\ .  \label{Tzzcs2}
\end{eqnarray}%
Writing $k^{2}e^{-lLk}=l^{-2}\partial _{L}^{2}e^{-lLk}$ and changing the
order of the differentiation and the integration over $k$, the integral is
expressed in terms of the function $K_{0}(lL\sqrt{m^{2}+\lambda ^{2}})$. By
using the integral representation (\ref{Kintergral}) for the latter we get
\begin{equation}
\langle T_{z}^{z}\rangle _{c}=-\frac{q}{2\pi ^{2}r^{2}}\partial
_{L}^{2}\sum_{l=1}^{\infty }\frac{\cos (2\pi l\tilde{\beta})}{l^{2}}%
\int_{0}^{\infty }dy\,e^{-y[1+l^{2}L^{2}/(2r^{2})]-m^{2}r^{2}/(2y)}\mathcal{J%
}(q,\alpha _{0},y).  \label{Tzz2}
\end{equation}%
After the differentiation one obtains%
\begin{eqnarray}
\langle T_{z}^{z}\rangle _{c} &=&\frac{q}{2\pi ^{2}r^{4}}\sum_{l=1}^{\infty
}\cos (2\pi l\tilde{\beta})\int_{0}^{\infty }dy\,y\left(
1-yl^{2}L^{2}/r^{2}\right)  \notag \\
&&\times e^{-y\left( 1+l^{2}L^{2}/(2r^{2})\right) -m^{2}r^{2}/(2y)}\mathcal{J%
}(q,\alpha _{0},y).  \label{Tzz3}
\end{eqnarray}%
By using the representation (\ref{Sum01}), after the integration over $y$,
we arrive at the final expression%
\begin{eqnarray}
\langle T_{z}^{z}\rangle _{c} &=&\frac{4m^{4}}{\pi ^{2}}\sum_{l=1}^{\infty
}\cos (2\pi l\tilde{\beta})\left[ \sideset{}{'}{\sum}_{k=0}^{p}(-1)^{k}\cos
(\pi k/q)\cos (2\pi k\alpha _{0})F_{z}^{(l)}(mL\sqrt{l^{2}+\rho _{k}^{2}}%
)\right.  \notag \\
&&+\left. \frac{q}{\pi }\int_{0}^{\infty }dx\frac{h(q,\alpha _{0},x)\sinh x}{%
\cosh (2qx)-\cos (q\pi )}F_{z}^{(l)}(mL\sqrt{l^{2}+\eta ^{2}(x)})\right] ,
\label{Tzz4}
\end{eqnarray}%
with the notation%
\begin{equation}
F_{z}^{(l)}(x)=f_{2}(x)-l^{2}L^{2}m^{2}f_{3}(x).  \label{Fzl}
\end{equation}%
Note that the relation $\langle T_{z}^{z}\rangle _{c}^{\mathrm{ren}}=\langle
T_{0}^{0}\rangle _{c}^{\mathrm{ren}}$ takes place only for a massless field.

\section{Properties of the energy-momentum tensor and the vacuum energy}

\label{sec4}

In this section we investigate the properties and the asymptotic behavior of
the VEVs found in previous section. We can check that the both contributions
to the VEV of the energy-momentum tensor obey the trace relation:%
\begin{equation}
\langle T_{\mu }^{\mu }\rangle _{s}^{\mathrm{ren}}=m\langle \bar{\psi}\psi
\rangle _{s}^{\mathrm{ren}},\;\langle T_{\mu }^{\mu }\rangle _{c}=m\langle
\bar{\psi}\psi \rangle _{c}.  \label{TrRel}
\end{equation}%
In particular, for a massless field the vacuum energy-momentum tensor is
traceless. Because the cosmic string spacetime is locally flat for $r>0$,
the trace anomaly is zero. Another important property obeyed by the VEV of
the energy-momentum tensor is its covariant conservation: $\nabla _{\mu
}\langle T_{\nu }^{\mu }\rangle =0$. For the problem under consideration
this equation is reduced to a single differential equation $\partial
_{r}(r\langle T_{r}^{r}\rangle )=\langle T_{\phi }^{\phi }\rangle $. We have
already proved this relation for separate straight cosmic string and
topological parts during the calculation of the corresponding VEVs in the
previous section (see (\ref{Tphir}) and (\ref{Tphirc})).

Combining the formulas obtained above, the part in the VEV of the
energy-momentum tensor corresponding to the geometry of a straight cosmic
string is presented as (no summation over $\mu $):%
\begin{eqnarray}
\langle T_{\mu }^{\mu }\rangle _{s}^{\mathrm{ren}} &=&\frac{2m^{4}}{\pi ^{2}}%
\left[ \sum_{k=1}^{p}(-1)^{k}{\cos }(\pi k/q){\cos }(2\pi k\alpha
_{0})F_{\mu }^{(0)}(2mrs_{k})\right.  \notag \\
&+&\left. \frac{q}{\pi }\int_{0}^{\infty }dx\,\frac{h(q,\alpha _{0},x)\sinh x%
}{\cosh (2qx)-\cos (q\pi )}F_{\mu }^{(0)}(2mr\cosh x)\right] ,  \label{Tmus}
\end{eqnarray}%
where $F_{\mu }^{(0)}(x)=f_{2}(x)$ for $\mu =0,r,z$, and the function $%
F_{\phi }^{(0)}(x)$ is defined by the relation (\ref{Fphi}). In the absence
of the magnetic flux along the string axis one has $\alpha _{0}=0$ and (\ref%
{Tmus}) is reduced to the expression given in \cite{Beze13b}. In this
special case an alternative integral representation for $\langle T_{\mu
}^{\mu }\rangle _{s}^{\mathrm{ren}}$ is derived in \cite{Tar08}. The case $%
q<2$, $\alpha _{0}=0$ was considered in \cite{BezeKh06}. For a massless
field and for $\alpha _{0}=0$, the renormalized VEV for the energy-momentum
tensor was found in \cite{Frol87,Dowk87}.

For the special case (\ref{gammmaSp}) with an integer $q$, the integral term
in (\ref{Tmus}) vanishes and we get simple expression%
\begin{equation}
\langle T_{\mu }^{\mu }\rangle _{s}^{\mathrm{ren}}=\frac{m^{4}}{\pi ^{2}}%
\sum_{k=1}^{q-1}{\cos }(\pi k/q)\cos ((2n+1)\pi k/q)F_{\mu }^{(0)}(2mrs_{k}).
\label{TmusSp}
\end{equation}%
Another simplification takes place for a massless field. In this case, by
using (\ref{fnuas}), from (\ref{Tmus}) we find (no summation over $\mu $)
\begin{equation}
\langle T_{\mu }^{\mu }\rangle _{s}^{\mathrm{ren}}=\delta _{\mu }\frac{%
s_{4}(q,\alpha _{0})}{4\pi ^{2}r^{4}},  \label{T00m0}
\end{equation}%
where%
\begin{equation}
\delta _{\mu }=1\text{ for }\mu =0,r,z,\;\delta _{\phi }=-3,  \label{delmu}
\end{equation}
and the function $s_{n}(q,\alpha _{0})$ is defined in (\ref{sq}). Note that,
by taking into account (\ref{s024}), for $\alpha _{0}=0$, Eq. (\ref{T00m0})
is reduced to the result given in \cite{Frol87,Dowk87}. For a massive field,
the expression in the right-hand side of (\ref{T00m0}) gives the leading
term in the asymptotic expansion of $\langle T_{0}^{0}\rangle _{s}^{\mathrm{%
ren}}$ for points near the string, $mr\ll 1$.

At large distance from the string, for $q>2$ the dominant contribution in (%
\ref{Tmus}) comes from the term $k=1$ and, in the leading order, we find:%
\begin{eqnarray}
\langle T_{0}^{0}\rangle _{s}^{\mathrm{ren}} &\approx &-\frac{%
m^{4}e^{-2mr\sin (\pi /q)}}{4\pi ^{3/2}}\frac{{\cos }(\pi /q){\cos }(2\pi
\alpha _{0})}{[mr\sin (\pi /q)]^{5/2}},  \notag \\
\langle T_{\phi }^{\phi }\rangle _{s}^{\mathrm{ren}} &\approx &-2mr\sin (\pi
/q)\langle T_{0}^{0}\rangle _{s}^{\mathrm{ren}} \ .  \label{TrenLarge}
\end{eqnarray}%
For $1\leqslant q\leqslant 2$ the sum over $k$ in the right-hand side of (%
\ref{Tmus}) is absent and the integral term is suppressed by the factor $%
e^{-2mr}$. In both cases $|\langle T_{0}^{0}\rangle _{s}^{\mathrm{ren}%
}|/|\langle T_{\phi }^{\phi }\rangle _{s}^{\mathrm{ren}}|\sim (mr)^{-1}\ll 1$%
.

\begin{figure}[tbph]
\begin{center}
\epsfig{figure=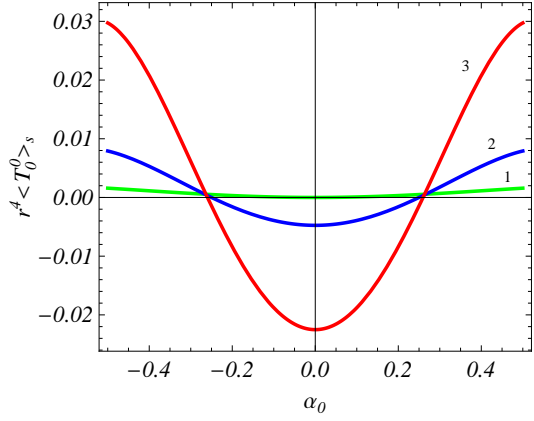,width=7.cm,height=5.5cm}
\end{center}
\caption{Quantity $r^{4}\langle T_{0}^{0}\rangle _{s}^{\mathrm{ren}}$ for a
massless fermionic field as a function of $\protect\alpha_0 $ for different
values of $q$ (numbers near the curves).}
\label{fig4}
\end{figure}

In figure \ref{fig4} we have plotted the energy density for a massless
fermionic field, multiplied by $r^{4}$, as a function of the parameters $%
\alpha_0 $ for different values of $q$ (numbers near the curves). As it is
seen, with dependence of the magnetic flux, the vacuum energy density can be
either positive or negative. In particular, the energy density is negative
in the absence of the magnetic flux. The maximal (positive) value of the
energy density is obtained for $\Phi _{\phi }=\Phi _{0}/2$. For some values
of $\alpha _{0}$ the VEV\ of the energy-momentum tensor vanishes. For these
values, the effects induced by the topology of the cosmic string spacetime
and by the magnetic flux compensate each other.

Now we turn to the investigation of the topological parts in the VEVs for
the components of the energy-momentum tensor. The $k=0$ terms in these parts
present the VEVs in Minkowski spacetime with spatial topology $R^{2}\times
S^{1}$ in the absence of the cosmic string and flux along the $z$-axis ($q=1$%
, $\alpha _{0}=0$). Denoting the corresponding VEV by $\langle T_{\mu }^{\mu
}\rangle _{c}^{(0)}$, from the formulas given above we have (no summation
over $\mu $):%
\begin{equation}
\langle T_{\mu }^{\mu }\rangle _{c}^{(0)}=\frac{2m^{4}}{\pi ^{2}}%
\sum_{l=1}^{\infty }{\cos }(2\pi l\tilde{\beta})F_{\mu }^{(l)}(mLl)\ ,
\label{Tc0}
\end{equation}%
with the functions
\begin{eqnarray}
F_{0}^{(l)}(x) &=&F_{r}^{(l)}(x)=f_{2}(x),  \notag \\
F_{\phi }^{(l)}(x) &=&\left( m^{2}L^{2}l^{2}-x^{2}\right) f_{3}(x)+f_{2}(x),
\label{Fphil1} \\
F_{z}^{(l)}(x) &=&f_{2}(x)-l^{2}L^{2}m^{2}f_{3}(x)\ .  \notag
\end{eqnarray}%
The corresponding results for a more general topology $R^{p}\times
(S^{1})^{D-p}$ are given in \cite{Bell09}. Now, combining the formulas
obtained before for separate components of the energy-momentum tensor, we
can see that the total VEV is written in the form (no summation over $\mu $)
\begin{eqnarray}
\langle T_{\mu }^{\mu }\rangle &=&\langle T_{\mu }^{\mu }\rangle _{c}^{(0)}+%
\frac{4m^{4}}{\pi ^{2}}\sideset{}{'}{\sum}_{l=0}^{\infty }{\cos }(2\pi l%
\tilde{\beta})  \notag \\
&&\times \left[ \sum_{k=1}^{p}(-1)^{k}\cos (\pi k/q)\cos (2\pi k\alpha
_{0})F_{\mu }^{(l)}(mL\sqrt{l^{2}+\rho _{k}^{2}})\right.  \notag \\
&& +\left. \frac{q}{\pi }\int_{0}^{\infty }dx\,\frac{h(q,\alpha _{0},x)\sinh
x}{\cosh (2qx)-\cos (q\pi )}F_{\mu }^{(l)}(mL\sqrt{l^{2}+\eta ^{2}(x)})%
\right] \ ,  \label{Tmutot}
\end{eqnarray}%
where the prime on the sum over $l$, as before, means that the term with $%
l=0 $ should be taken with the coefficient 1/2. The latter coincides with $%
\langle T_{\mu }^{\mu }\rangle _{s}$ and the $l\neq 0$ terms come from the
compactification of the string along its axis. The VEV (\ref{Tmutot}) is an
even periodic function of $\tilde{\beta}$ and $\alpha _{0}$ with the period
equal to 1. It is symmetric under the replacement $\tilde{\beta}\rightarrow
1-\tilde{\beta}$.

Simpler expressions for the VEVs are obtained in two particular cases. For
integer values of the parameter $q$ and for $\alpha _{0}$ given by (\ref%
{gammmaSp}) one has $h(q,\alpha _{0},x)=0$ and (\ref{Tmutot}) is reduced to
\begin{equation}
\langle T_{\mu }^{\mu }\rangle =\langle T_{\mu }^{\mu }\rangle _{c}^{(0)}+%
\frac{2m^{4}}{\pi ^{2}}\sum_{k=1}^{q-1}\cos (\pi k/q)\cos ((2n+1)\pi k/q)%
\sideset{}{'}{\sum}_{l=0}^{\infty }{\cos }(2\pi l\tilde{\beta})F_{\mu
}^{(l)}(mL\sqrt{l^{2}+\rho _{k}^{2}})\ .  \label{TmutotSp}
\end{equation}%
For a massless field, by using the asymptotic expression (\ref{fnuas}), one
gets (no summation over $\mu $)%
\begin{eqnarray}
\langle T_{\mu }^{\mu }\rangle &=&\langle T_{\mu }^{\mu }\rangle _{c}^{(0)}+%
\frac{8\delta _{\mu }}{\pi ^{2}L^{4}}\left[ \sum_{k=1}^{p}(-1)^{k}\cos (\pi
k/q)\cos (2\pi k\alpha _{0})C(\tilde{\beta},\rho _{k})\right.  \notag \\
&& + \left. \frac{q}{\pi }\int_{0}^{\infty }dx\,\frac{h(q,\alpha _{0},x)C(%
\tilde{\beta},\eta (x))\sinh x}{\cosh (2qx)-\cos (q\pi )}\right] \ ,
\label{Tmutotm0}
\end{eqnarray}%
where $\delta _{\mu }$ is given by (\ref{delmu}) and we have defined the
function%
\begin{equation}
C(\tilde{\beta},x)=\sideset{}{'}{\sum}_{l=0}^{\infty }\frac{\cos (2\pi l%
\tilde{\beta})}{\left( l^{2}+x^{2}\right) ^{2}} \ .  \label{Cbet}
\end{equation}%
For the series (\ref{Cbet}) one has \cite{Prud86}%
\begin{eqnarray}
C(\tilde{\beta},x) &=&\frac{\pi ^{2}\cosh (2\pi \tilde{\beta}x)}{4x^{2}\sinh
^{2}(\pi x)}  \notag \\
&&+\pi \frac{\cosh [\pi (1-2\tilde{\beta})x]+2\pi \tilde{\beta}x\sinh [\pi
(1-2\tilde{\beta})x]}{4x^{3}\sinh (\pi x)} \ ,  \label{Cbet1}
\end{eqnarray}%
with $0\leqslant \tilde{\beta}\leqslant 1$. In this case, the Minkowskian
part reads:%
\begin{eqnarray}
\langle T_{\mu }^{\mu }\rangle _{c}^{(0)} &=&\frac{4\delta _{\mu }}{\pi
^{2}L^{4}}\lim_{x\rightarrow 0}\left[ C(\tilde{\beta},x)-\frac{1}{2x^{4}}%
\right]  \notag \\
&=&\frac{4\pi ^{2}\delta _{\mu }}{3L^{4}}\left[ \frac{1}{30}-\tilde{\beta}%
^{2}(1-\tilde{\beta})^{2}\right] \ .  \label{Tc0m0}
\end{eqnarray}

Now we consider the asymptotics of the topological part in the VEV of the
energy-momentum tensor near the string and at large distances. The
topological part is finite on the string:%
\begin{equation}
\langle T_{\mu }^{\mu }\rangle _{c}|_{r=0}=[1+2s_{0}(q,\alpha _{0})]\langle
T_{\mu }^{\mu }\rangle _{c}^{(0)}\ ,  \label{TmucStr}
\end{equation}%
where $\langle T_{\mu }^{\mu }\rangle _{c}^{(0)}$ is given by expressions ( %
\ref{Tc0}) and (\ref{Tc0m0}) for massive and massless fields respectively.
In the absence of magnetic flux along the string axis, $\alpha _{0}=0$, and
in accordance with (\ref{s024}) the topological part vanishes on the string.
The part $\langle T_{\mu }^{\mu }\rangle _{s}$ diverge on the string and,
hence, it dominates in the total VEV for points near the string, $r\ll L$.
In the opposite limit, $r\gg L$, and for a massless field, in (\ref{Tmutotm0}%
) the argument of the function $C(\tilde{\beta},x)$ is large, $x\gg 1$. In
this case we use the asymptotic formula $C(\tilde{\beta},x)\approx \pi
^{2}\sigma e^{-2\pi \sigma x}/(2x^{2})$, with $\sigma =\min (\tilde{\beta},1-%
\tilde{\beta})$ and $0<\tilde{\beta}<1$. For $\tilde{\beta}=0$ one has $%
C(0,x)\approx \pi /(4x^{3})$, $x\gg 1$. For $q>2$ and $0<\tilde{\beta}<1$,
the dominant contribution to the second term in the right-hand side of (\ref%
{Tmutotm0}) comes from the $k=1$ term and we have
\begin{equation}
\langle T_{\mu }^{\mu }\rangle \approx \langle T_{\mu }^{\mu }\rangle
_{c}^{(0)}-\frac{\delta _{\mu }\sigma \cos (\pi /q)\cos (2\pi \alpha _{0})}{%
L^{2}r^{2}\sin ^{2}(\pi /q)}e^{-4\pi \sigma r\sin (\pi /q)/L} \ .
\label{TmuLarge}
\end{equation}%
For $q\leqslant 2$, in (\ref{Tmutotm0}) the sum over $k$ is absent and the
contribution of the integral term is suppressed by the factor $e^{-4\pi
\sigma r/L}$. Hence, in the case $q\leqslant 2$ the suppression of the
effects induced by the string at large distances are stronger. For $\tilde{%
\beta}=0$ we have the asymptotic formula
\begin{equation}
\langle T_{\mu }^{\mu }\rangle \approx \langle T_{\mu }^{\mu }\rangle
_{c}^{(0)}+\frac{\delta _{\mu }s_{3}(q,\alpha _{0})}{4\pi Lr^{3}} \ ,
\label{TmuLargeb}
\end{equation}%
with the function $s_{n}(q,\alpha _{0})$ defined in (\ref{sq}).

\begin{figure}[tbph]
\begin{center}
\epsfig{figure=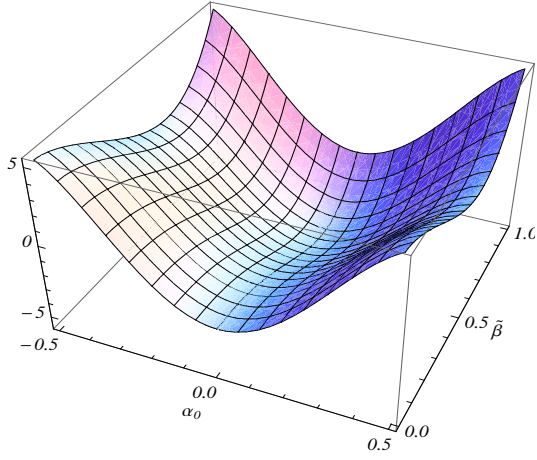,width=7.cm,height=6.cm}
\end{center}
\caption{Energy density, $L^{4}\langle T_{0}^{0}\rangle $, for a massless
fermionic field as a function of $\protect\alpha _{0}$ and $\tilde{\protect%
\beta}$ for fixed values $q=2.5$ and $r/L=0.25$.}
\label{fig5}
\end{figure}

\begin{figure}[tbph]
\begin{center}
\epsfig{figure=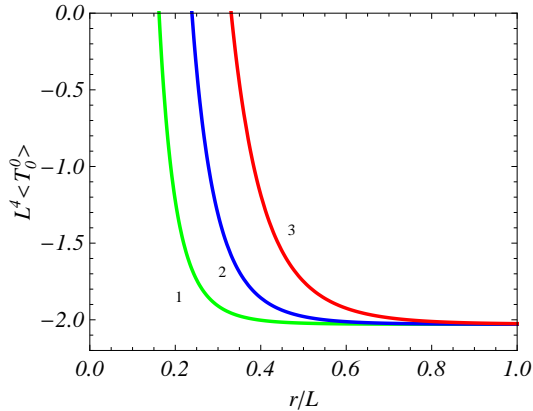,width=7.cm,height=5.5cm}
\end{center}
\caption{Energy density, $L^{4}\langle T_{0}^{0}\rangle $, for a massless
fermionic field as a function of the distance from the string for $\protect%
\alpha _{0}=\tilde{\protect\beta}=0.5$. The numbers near the curves
correspond to the value of the parameter $q$.}
\label{fig6}
\end{figure}

In figure \ref{fig5} we have presented the quantity $L^{4}\langle
T_{0}^{0}\rangle $ for a massless fermionic field as a function of the
parameters $\alpha _{0}$ and $\tilde{\beta}$ for fixed values $q=2.5$ and $%
r/L=0.25$. The dependence of the same function on the distance from the
string is displayed in figure \ref{fig6} for $\alpha _{0}=\tilde{\beta}=0.5$%
. The numbers near the curves correspond to the value of the parameter $q$.

From the discussion given above it follows that we can decompose the VEV of
the energy density as%
\begin{equation}
\langle T_{0}^{0}\rangle =\langle T_{0}^{0}\rangle _{c}^{(0)}+\langle
T_{0}^{0}\rangle _{s}^{\mathrm{ren}}+\langle T_{0}^{0}\rangle _{c}^{(s)},
\label{T00dec}
\end{equation}%
where the last term is the contribution of the cosmic string to the
topological part to the energy density. The distribution of the energy
density $\langle T_{0}^{0}\rangle _{c}^{(0)}$, corresponding to the
Minkowski spacetime compactified along the $z$-direction, is uniform. The
part $\langle T_{0}^{0}\rangle _{s}^{\mathrm{ren}}$ behaves near the string
as $1/r^{4}$ and, hence, the corresponding contribution to the vacuum energy
is divergent. We can evaluate the total vacuum energy (per unit length of
the string) induced by the straight string in the region $r\geqslant r_{0}>0$%
:%
\begin{equation}
E_{s,r\geqslant r_{0}}=\phi _{0}\int_{r_{0}}^{\infty }dr\,r\langle
T_{0}^{0}\rangle _{s}^{\mathrm{ren}}.  \label{Es}
\end{equation}%
By using the formula $\int_{a}^{\infty }dx\,xf_{\nu }(x)=f_{\nu -1}(a)$ and (%
\ref{Tmus}), one gets%
\begin{eqnarray}
E_{s,r\geqslant r_{0}} &=&\frac{m^{2}}{\pi ^{2}}\left[ \frac{\pi }{q}%
\sum_{k=1}^{p}\frac{(-1)^{k}}{s_{k}^{2}}{\cos }(\pi k/q){\cos }(2\pi k\alpha
_{0})f_{1}(2mr_{0}s_{k})\right.  \notag \\
&+&\left. \int_{0}^{\infty }dx\frac{h(q,\alpha _{0},x)\sinh x\,}{\cosh
(2qx)-\cos (q\pi )}\frac{f_{1}(2mr_{0}\cosh x)}{\cosh ^{2}x}\right] .
\label{Es1}
\end{eqnarray}%
For a massless field this expression is written as:
\begin{equation}
E_{s,r\geqslant r_{0}}=\frac{s_{4}(q,\alpha _{0})}{4\pi ^{2}r_{0}^{2}} \ .
\label{Es1m0}
\end{equation}%
In the absence of the magnetic flux $\alpha _{0}=0$, this formula is further
simplified by using (\ref{s024}).

The last term in (\ref{T00dec}) is given by the part in (\ref{Tmutot}) with $%
l\neq 0$. For the corresponding contribution to the total vacuum energy one
has%
\begin{equation}
E_{c}^{s}=\int_{0}^{L}dz\int_{0}^{\phi _{0}}d\phi \int_{0}^{\infty
}dr\,r\langle T_{0}^{0}\rangle _{c}^{(s)}=\frac{2s_{2}(q,\alpha _{0})}{\pi qL%
}F(\tilde{\beta},mL)\ ,  \label{Esc}
\end{equation}%
where the function $s_{n}(q,\alpha _{0})$ is given by the expression (\ref%
{sq}) and we have defined the function%
\begin{equation}
F(\tilde{\beta},x)=x\sum_{l=1}^{\infty }\frac{{\cos }(2\pi l\tilde{\beta})}{l%
}K_{1}(lx)\ .  \label{Fbetx}
\end{equation}%
For a massless field the expression for $E_{c}^{s}$ is obtained from (\ref%
{Esc}), by taking into account that $F(\tilde{\beta},0)=\pi ^{2}(1/6+\tilde{%
\beta}^{2}-\tilde{\beta})$. In the absence of the magnetic flux along the
axis of the string one has $\alpha _{0}=0$ and, by using (\ref{s024}), one
finds%
\begin{equation}
E_{c}^{s}=\frac{1-q^{2}}{6\pi qL}F(\tilde{\beta},mL)\ .  \label{Esc2}
\end{equation}%
In particular, in the absence of magnetic flux enclosed by the compact
dimension, $\Phi _{z}=0$, this energy is negative for an untwisted fermionic
field ($\beta =0$)\ and is positive for twisted field ($\beta =1/2$).

\section{Conclusion}

\label{conc}

In this paper we have investigated the FC and the VEV of the energy-momentum
tensor for a charged massive fermion field in the compactified cosmic string
spacetime considering the presence of a constant vector potential. Though
the magnetic field strength is zero, the nontrivial spatial topology gives
arise to Aharonov-Bohm-like effects on the VEVs. We have assumed that the
field operator obeys a quasiperiodicity condition along the $z$-axis with
the period $L$, exhibited in (\ref{Period}). The phase in this condition and
the component of the gauge field along the string axis are related through a
gauge transformation and the VEVs depend on the combination (\ref{bett}). By
applying the Abel-Plana summation formula to the series over the quantum
number corresponding to the compact dimension, we have explicitly extracted
from the VEVs the contributions corresponding to the geometry of a straight
cosmic string. The compactification does not change the local geometry, so
the renormalization of the VEVs is reduced to the one for the straight
cosmic string spacetime part. Outside the string core the background
geometry is flat and for the renormalization we need just subtract the parts
corresponding to topologically trivial Minkowski spacetime in the absence of
magnetic fluxes. The application of the formula (\ref{Sum01}) allowed us to
separate explicitly these parts. In this way, we have provided closed
analytic expressions for the both renormalized straight cosmic string and
topological parts as functions of the planar angle deficit and two kinds of
magnetic fluxes: running along the string axis and enclosed by the compact
dimension. In particular, both FC and the vacuum energy-momentum tensor, are
even periodic functions of these fluxes with the period equal to the flux
quantum.

The renormalized FC in the geometry of a straight cosmic string is given by
the expression (\ref{FC4}). With respect to the dependence on the magnetic
flux running along the string axis, the FC can be either positive or
negative for a massive field and it vanishes for a massless field. In
particular, in the absence of the magnetic flux the FC is positive. The
straight cosmic string part in the FC diverges on the string as $1/r^{2}$.
At distances from the string larger than the Compton wavelength of the
fermionic particle, it decays as $e^{-2mr}$ for $1\leqslant q\leqslant 2$
and as $e^{-2mr\sin (\pi /q)}$ for $q>2$. The part in the FC induced by the
compactification of the cosmic string axis is given by (\ref{FCc6}) (an
alternative expression is provided by (\ref{FCc5})). The $k=0$ term in this
expression, see (\ref{FCM}), corresponds to the FC in Minkowski spacetime
with spatial topology $R^{2}\times S^{1}$, denoted as $\langle \bar{\psi}%
\psi \rangle _{c}^{(0)}$. The topological part is finite on the string and
vanishes in the absence of the magnetic flux along the string axis. For
points near the string, the dominant contribution into the total FC comes
from the straight cosmic string part, $\langle \bar{\psi}\psi \rangle _{s}^{%
\mathrm{ren}}$. On the other hand, at large distances from it the effects
induced by the string decay exponentially and the FC tends to the limiting
value which coincides with $\langle \bar{\psi}\psi \rangle _{c}^{(0)}$. For
small values of the length of compact dimension, the asymptotic behavior of
the topological part in the FC crucially depends on the values of the
parameter $\tilde{\beta}$ describing the magnetic flux enclosed by the
compact dimension, i.e. whether it is zero or not. For $0<\tilde{\beta}<1$,
the effects of the cosmic string in the topological part of the FC are
suppressed by the factor $\exp [-4\pi \tilde{\beta}r\sin (\pi /q)/L]$ for $%
q>2$ and by the factor $\exp (-4\pi \tilde{\beta}r/L)$ for $1\leqslant
q\leqslant 2$. In the case $\tilde{\beta}=0$ and for small values of $L$,
the asymptotic expression for the FC is given by (\ref{FCsmalL}), where $%
\langle \bar{\psi}\psi \rangle _{s}^{(D=2)}$ is the FC in a $(2+1)$%
-dimensional conical spacetime. The numerical calculations have shown that,
at a given point, both the sign and the absolute value of the FC can be
effectively tuned by changing the magnetic fluxes.

Another important characteristic of the fermionic vacuum is the VEV of the
energy-momentum tensor. The main steps of the evaluation for the separate
components of this tensor are presented in Section \ref{sec3}. Combined
expressions for both straight cosmic string and topological parts, and their
properties are presented in Section \ref{sec4}; in this way we have
explicitly shown that these parts obey, separately, the trace relation (\ref%
{TrRel}) and covariant conservation equation. In particular, the vacuum
energy-momentum tensor is traceless for a massless fermionic field. The
renormalized VEV of the energy-momentum tensor in the geometry of a straight
cosmic string is given in a general form by (\ref{Tmus}). This expression
includes various special cases previously described in the literature. The
radial and axial stresses are equal to the energy density. For the axial
stress this result could be directly obtained from the boost invariance
along the string axis. For a massless field the general formula is
simplified to (\ref{T00m0}). In the absence of the magnetic flux we recover
the well-known result derived in \cite{Frol87,Dowk87}. In the general case
of a massive field, near the string the straight cosmic string part behaves
as $1/r^{4}$, and it dominates in the total VEV of the energy-momentum
tensor. At large distances from the string, $mr\gg 1$, the leading terms in
the corresponding asymptotic expansions are given by (\ref{TrenLarge}). Note
that in this limit $|\langle T_{0}^{0}\rangle _{s}^{\mathrm{ren}}|\ll
|\langle T_{\phi }^{\phi }\rangle _{s}^{\mathrm{ren}}|$. As it is
illustrated in figure \ref{fig4}, with a dependence on the magnetic flux,
the vacuum energy density in the geometry of a straight cosmic string can be
either positive or negative. In particular, the energy density is negative
in the absence of the magnetic flux.

In the compactified cosmic string spacetime, the total VEV of the
energy-momentum tensor is given by (\ref{Tmutot}). The term $l=0$ in this
expression corresponds to the VEV for the geometry of a straight cosmic
string and the remaining part is induced by the compactification of the
string axis. The term $\langle T_{\mu }^{\mu }\rangle _{c}^{(0)}$, defined
in (\ref{Tc0}), presents the VEV in Minkowski spacetime with spatial
topology $R^{2}\times S^{1}$ in the absence of the cosmic string and flux
along the $z$-axis. Alternative expressions for the topological parts in the
VEVs are given by (\ref{T00c2c}), (\ref{Tppc3}) and (\ref{Tzz3}). As in the
case of the straight cosmic string geometry, the radial stress is equal to
the energy density. However, the compactification breaks the boost
invariance along the string axis and the axial stress differs from the
energy density. Simpler expressions are obtained for two special cases. For
integer values of the parameter $q$ and for $\alpha _{0}$ given by (\ref%
{gammmaSp}), the integral term vanishes and we get the expression (\ref%
{TmutotSp}). For a massless field the VEV of the energy-momentum tensor is
reduced to (\ref{Tmutotm0}). In this special case, the axial stress is equal
to the energy density. The topological part in the VEV of the
energy-momentum tensor is finite on the string with the limiting value given
by (\ref{TmucStr}). In the absence of the magnetic flux along the string
axis the topological part vanishes on the string. At large distances from
the string and for a massless field, the part in the VEV of the
energy-momentum tensor induced by the string is suppressed by the factor $%
e^{-4\pi \sigma r\sin (\pi /q)/L}$ (see (\ref{TmuLarge})) for $q>2$ and by $%
e^{-4\pi \sigma r/L}$ for $q\leqslant 2$ with $0<\tilde{\beta}<1$ and $%
\sigma =\min (\tilde{\beta},1-\tilde{\beta})$. For $\tilde{\beta}=0$ we have
a power-law decay given by (\ref{TmuLargeb}).

In addition to the local characteristics of the vacuum state, we have also
evaluated the part in the topological Casimir energy induced by the string.
The latter is given by a simple formula (\ref{Esc}). In the absence of the
magnetic flux along the axis of the string this formula is further
simplified to (\ref{Esc2}). If, in addition, there is no magnetic flux
enclosed by the compact dimension, the corresponding energy is
negative/positive for untwisted/twisted fermionic fields. The sign of the
vacuum energy can be controlled by tuning the values of the magnetic fluxes.

\section*{Acknowledgments}

ERBM thanks Conselho Nacional de Desenvolvimento Cient\'{\i}fico e Tecnol%
\'{o}gico (CNPq) for partial financial support. AAS was supported by State
Committee Science MES RA, within the frame of the research project No. SCS
13-1C040.


\begin{thebibliography}{99}
\bibitem{Eliz94} E. Elizalde, S.D. Odintsov, A. Romeo, A.A. Bytsenko and S.
Zerbini, \textit{Zeta regularization techniques with applications} (World
Scientific, Singapore, 1994).

\bibitem{Most} V.M. Mostepanenko and N.N. Trunov, \textit{The Casimir Effect
and Its Applications} (Clarendon, Oxford, 1997).

\bibitem{Milton} K.A. Milton, \textit{The Casimir Effect: Physical
Manifestation of Zero-Point Energy} (World Scientific, Singapore, 2002).

\bibitem{Bord} M. Bordag, G.L. Klimchitskaya, U. Mohideen, and V.M.
Mostepanenko, \textit{Advances in the Casimir Effect} (Oxford University
Press, Oxford, 2009); \textit{Lecture Notes in Physics: Casimir Physics},
Vol. 834, edited by D. Dalvit, P. Milonni, D. Roberts, and F. da Rosa
(Springer, Berlin, 2011).

\bibitem{Eliz01} E. Elizalde, Phys. Lett. B \textbf{516}, 143 (2001).

\bibitem{Gard} C.L. Gardner, Phys. Lett. B \textbf{524}, 21 (2002); K.A.
Milton, Grav. Cosmol. \textbf{9}, 66 (2003).

\bibitem{ASah} A.A. Saharian, Phys. Rev. D \textbf{70}, 064026 (2004).

\bibitem{Eliz06} E. Elizalde, J. Phys. A \textbf{39}, 6299 (2006); A.A.
Saharian, Phys. Rev. D \textbf{74}, 124009 (2006).

\bibitem{Gree} B. Green and J. Levin, J. High Energy Phys. \textbf{11}, 096
(2007).

\bibitem{Bur} P. Burikham, A. Chatrabhuti, P. Patcharamaneepakorn, and K.
Pimsamarn, J. High Energy Phys. \textbf{07}, 013 (2008).

\bibitem{Zhi} A.R. Zhitnitsky, Phys. Rev. D \textbf{86}, 045026 (2012).

\bibitem{Bell09} S. Bellucci and A.A. Saharian, Phys. Rev. D \textbf{79},
085019 (2009).

\bibitem{Eliz11} E. Elizalde, S.D. Odintsov, and A.A. Saharian, Phys. Rev. D
\textbf{83}, 105023 (2011).

\bibitem{Saha13} E. R. Bezerra de Mello and A. A. Saharian, Eur. Phys. J. C.
\textbf{73}, 2532 (2013).

\bibitem{Mello12} E.R. Bezerra de Mello and A.A. Saharian, Class. Quantum
Grav. \textbf{29}, 035006 (2012).

\bibitem{V-S} T.W.B. Kibble, Phys. Rep. \textbf{67}, 183 (1980); A. Vilenkin
and E.P.S. Shellard, \textit{Cosmic Strings and Other Topological Defects}
(Cambridge University Press, Cambridge, England, 1994).

\bibitem{Berezinski} V. Berezinski, B. Hnatyk and A. Vilenkin, Phys. Rev. D
\textbf{64}, 043004 (2001).

\bibitem{Damour} T. Damour and A. Vilenkin, Phys. Rev. Lett. \textbf{85},
3761 (2000).

\bibitem{Bhattacharjee} P. Bhattacharjee and G. Sigl, Phys. Rep. \textbf{327}%
, 109 (2000).

\bibitem{Sarangi} S. Sarangi and S.-H. Henry Tye, Phys. Lett. B \textbf{536}%
, 185 (2002).

\bibitem{Copeland} E.J. Copeland, R.C. Myers, and J. Polchinski, J. High
Energy Phys. \textbf{06}, 013 (2004).

\bibitem{Dvali} G. Dvali and A. Vilenkin, J. Cosmol. Astropart. Phys.
\textbf{03}, 010 (2004).

\bibitem{Kibb04} T.W.B. Kibble, arXiv:atro-ph/0410073.

\bibitem{Hell86} T.M. Helliwell and D.A. Konkowski, Phys. Rev. D \textbf{34}
, 1918 (1986).

\bibitem{Line87} B. Linet, Phys. Rev. D \textbf{35}, 536 (1987).

\bibitem{Frol87} V.P. Frolov and E.M. Serebriany, Phys. Rev. D \textbf{35},
3779 (1987).

\bibitem{Dowk87} J.S. Dowker, Phys. Rev. D \textbf{36}, 3095 (1987); J.S.
Dowker, Phys. Rev. D \textbf{36}, 3742 (1987).

\bibitem{Davi88} P.C.W. Davies and V. Sahni, Class. Quantum Grav. \textbf{5}
, 1 (1988).

\bibitem{Smit89} A.G. Smith, in \textit{The Formation and Evolution of
Cosmic Strings}, Proceedings of the Cambridge Workshop, Cambridge, England,
1989, edited by G.W. Gibbons, S.W. Hawking, and T. Vachaspati (Cambridge
University Press, Cambridge, England, 1990).

\bibitem{Alle90} B. Allen and A.C. Ottewill, Phys. Rev. D \textbf{42}, 2669
(1990); B. Allen, J.G. Mc Laughlin, and A.C. Ottewill, Phys. Rev. D \textbf{%
45}, 4486 (1992); B. Allen, B.S. Kay, and A.C. Ottewill, Phys. Rev. D
\textbf{53}, 6829 (1996).

\bibitem{Sour92} T. Souradeep and V. Sahni, Phys. Rev. D \textbf{46}, 1616
(1992).

\bibitem{Shir92} K. Shiraishi and S. Hirenzaki, Class. Quantum Grav. \textbf{%
9}, 2277 (1992).

\bibitem{Beze94} V.B. Bezerra and E.R. Bezerra de Mello, Class. Quantum
Grav. \textbf{11}, 457 (1994); E.R. Bezerra de Mello, Class. Quantum Grav.
\textbf{11}, 1415 (1994).

\bibitem{Cogn94} G. Cognola, K. Kirsten, and L. Vanzo, Phys. Rev. D \textbf{%
49}, 1029 (1994).

\bibitem{More95} E.S. Moreira Jnr, Nucl. Phys. B \textbf{451}, 365 (1995).

\bibitem{Iell97} D. Iellici, Class. Quantum Grav. \textbf{14}, 3287 (1997).

\bibitem{Khus99} N.R. Khusnutdinov and M. Bordag, Phys. Rev. D \textbf{59},
064017 (1999).

\bibitem{BezeKh06} V.B. Bezerra and N.R. Khusnutdinov, Class. Quantum Grav.
\textbf{23}, 3449 (2006).

\bibitem{Bard10} V.M. Bardeghyan and A.A. Saharian, J. Contemp. Phys. (Arm.
Acad. Sci.) \textbf{45}, 1 (2010); A.A. Saharian and A.S. Kotanjyan, Eur.
Phys. J. C \textbf{71}, 1765 (2011); E.R. Bezerra de Mello, V.B. Bezerra,
H.F. Mota, and A.A. Saharian, Phys. Rev. D \textbf{86}, 065023 (2012).

\bibitem{charged} J.S. Dowker, Phys. Rev. D \textbf{36}, 3742 (1987).

\bibitem{charged1} M.E.X. Guimar\~{a}es and B. Linet, Commun. Math. Phys.
\textbf{165}, 297 (1994).

\bibitem{charged3} J. Spinelly and E.R. Bezerra de Mello, Class. Quantum
Grav. \textbf{20} 874, (2003); J. Spinelly and E.R. Bezerra de Mello, Int.
J. Mod. Phys. A, \textbf{17}, 4375 (2002).

\bibitem{Spin} J. Spinelly and E.R. Bezerra de Mello, Int. J. Mod. Phys. D
\textbf{13}, 607 (2004); J. Spinelly and E.R. Bezerra de Mello, Nucl Phys. B
(Proc. Suppl.) \textbf{127}, 77 (2004).

\bibitem{Spin08} J. Spinelly and E. R. Bezerra de Mello, JHEP \textbf{09},
005 (2008).

\bibitem{Sira} L. Sriramkumar, Class. Quantum Grav. \textbf{18}, 1015 (2001).

\bibitem{Yu} Yu.A. Sitenko and N.D. Vlasii, Class. Quantum Grav. \textbf{26}%
, 195009 (2009).

\bibitem{Mello10} E. R. Bezerra de Mello, Class. Quantum Grav. \textbf{27},
095017 (2010).

\bibitem{Saha10} E. R. Bezerra de Mello, V.B. Bezerra, A.A. Saharian, and
V.M. Bardeghyan, Phys. Rev. D \textbf{82}, 085033 (2010).

\bibitem{Brev95} I. Brevik and T. Toverud, Class. Quantum Grav. \textbf{12},
1229 (1995).

\bibitem{Tar06} E.R. Bezerra de Mello, V.B. Bezerra, A.A. Saharian, and A.S.
Tarloyan, Phys. Rev. D \textbf{74}, 025017 (2006).

\bibitem{Bezerra07} E.R. Bezerra de Mello, V.B. Bezerra, and A.A. Saharian,
Phys. Lett. B \textbf{645}, 245 (2007).

\bibitem{Tar08} E.R. Bezerra de Mello, V.B. Bezerra, A.A. Saharian, and A.S.
Tarloyan, Phys. Rev. D \textbf{78}, 105007 (2008).

\bibitem{Fucci11} G. Fucci and K. Kirsten, JHEP \textbf{1103}, 016 (2011).

\bibitem{EA} E.R. Bezerra de Mello and A.A. Saharian, Class. Quantum Grav.
\textbf{28}, 145008 (2011).

\bibitem{Kirs11} G. Fucci and K. Kirsten, J. Phys. A \textbf{44}, 295403
(2011).

\bibitem{Bell11} S. Bellucci, E.R. Bezerra de Mello, and A.A. Saharian,
Phys. Rev. D \textbf{83}, 085017 (2011).

\bibitem{Nest11} V.V. Nesterenko and I.G. Pirozhenko, Class. Quantum Grav.
\textbf{28}, 175020 (2011)

\bibitem{Moraes} E.R. Bezerra de Mello, F. Moraes, and A.A. Saharian, Phys.
Rev. D \textbf{85}, 045016 (2012).

\bibitem{Gri12} E.R. Bezerra de Mello, A.A. Saharian, and A.Kh. Grigoryan,
J. Phys. A: Math. Theor. \textbf{45}, 374011 (2012).

\bibitem{Beze13b} E.R. Bezerra de Mell, A.A. Saharian, and S.V. Abajyan,
Class. Quantum Grav. \textbf{30}, 015002 (2013).

\bibitem{Bell10} S. Bellucci, A.A. Saharian, and V.M. Bardeghyan, Phys. Rev.
D \textbf{82}, 065011 (2010).

\bibitem{SahaBook} A. A. Saharian, \textit{The Generalized Abel-Plana
Formula with Applications to Bessel Functions and Casimir Effect} (Yerevan
State University Publishing House, Yerevan, 2008); Report No. ICTP/2007/082;
arXiv:0708.1187.

\bibitem{Grad} I.S. Gradshteyn and I.M. Ryzhik, \textit{Table of Integrals,
Series and Products} (Academic Press, New York, 1980).

\bibitem{Prud86} A.P. Prudnikov, Yu.A. Brychkov, and O.I. Marichev, \textit{%
Integrals and Series} (Gordon and Breach, New York, 1986), Vol. 1.
\end{thebibliography}
\end{document}